%% file: main.tex
\documentclass[lettersize, journal]{IEEEtran}
\ifCLASSINFOpdf
\else
\fi
\usepackage{url}
\usepackage{hyperref}
\usepackage{graphicx}
\usepackage{algorithmic}
\usepackage{subfigure}
\usepackage{diagbox}
\usepackage{amsmath}
\usepackage{xcolor}
\usepackage{booktabs}
\usepackage{multirow}
\usepackage{balance}
\usepackage{listings}
\usepackage{booktabs}
\usepackage{pifont}
\usepackage{enumitem}
\usepackage[ruled,vlined,linesnumbered]{algorithm2e}
\usepackage[inkscapelatex=false]{svg}
\usepackage{bm}
\usepackage{array}
\usepackage{ragged2e}
\usepackage{tabularray}
\usepackage[normalem]{ulem}
\usepackage{enumitem}
\usepackage{amsfonts,amssymb} 
\usepackage{blindtext}
\usepackage{makecell}

\newcommand{\etal}{\textit{et al}. }
\newcommand{\ie}{\textit{i}.\textit{e}.}
\newcommand{\eg}{\textit{e}.\textit{g}.}
\newcommand{\etc}{\textit{etc}}

\usepackage{tikz}
\newcommand*\emptycirc[1][1ex]{\tikz\draw (0,0) circle (#1);} 
\newcommand*\halfcirc[1][1ex]{%
	\begin{tikzpicture}
	\draw[fill] (0,0)-- (90:#1) arc (90:270:#1) -- cycle ;
	\draw (0,0) circle (#1);
	\end{tikzpicture}}
\newcommand*\fullcirc[1][1ex]{\tikz\fill (0,0) circle (#1);}

\begin{document}

\title{Towards Privacy-Preserving and Personalized Smart Homes via Tailored Small Language Models}

\author{
    Xinyu Huang,~\IEEEmembership{Student Member,~IEEE,}
    Leming Shen, \IEEEmembership{Student Member, IEEE,}
    Zijing Ma, \IEEEmembership{Student Member, IEEE,}
    Yuanqing Zheng, \IEEEmembership{Senior Member, IEEE,}
\IEEEcompsocitemizethanks{
    \IEEEcompsocthanksitem X. Huang, L. Shen, Z. Ma, Y. Zheng are with the Department of Computing, The Hong Kong Polytechnic University, Hong Kong SAR, China. E-mail: unixy-xinyu.huang@connect.polyu.hk, leming.shen@connect.polyu.hk, zijing.ma@connect.polyu.hk, and csyqzheng@comp.polyu.edu.hk.
}
    \thanks{Corresponding author: Yuanqing Zheng.}
}

\markboth{IEEE TRANSACTIONS ON MOBILE COMPUTING,~Vol.~X, No.~X, July~2025}%
{Huang \MakeLowercase{\textit{et al.}}: Towards Privacy-Preserving and Personalized Smart Home Serving}

\maketitle

\begin{abstract}
Large Language Models (LLMs) have showcased remarkable generalizability in language comprehension and hold significant potential to revolutionize human-computer interaction in smart homes. Existing LLM-based smart home assistants typically transmit user commands, along with user profiles and home configurations, to remote servers to obtain personalized services. However, users are increasingly concerned about the potential privacy leaks to the remote servers. To address this issue, we develop \textit{HomeLLaMA}, an on-device assistant for privacy-preserving and personalized smart home serving with a tailored small language model (SLM). \textit{HomeLLaMA} learns from cloud LLMs to deliver satisfactory responses and enable user-friendly interactions. Once deployed, \textit{HomeLLaMA} facilitates proactive interactions by continuously updating local SLMs and user profiles. To further enhance user experience while protecting their privacy, we develop \textit{PrivShield} to offer an optional privacy-preserving LLM-based smart home serving for those users, who are unsatisfied with local responses and willing to send less-sensitive queries to remote servers. For evaluation, we build a comprehensive benchmark \textit{DevFinder} to assess the service quality. Extensive experiments and user studies ($M=100$) demonstrate that \textit{HomeLLaMA} can provide personalized services while significantly enhancing user privacy.
\end{abstract}


\section{Introduction}
\label{sec:introduction}
\input{sections/01_introduction}

\section{Related Work}
\input{sections/10_related_work}

\section{Motivation and Challenges}
\input{sections/02_motivation}

\section{Design of HomeLLaMA}
\input{sections/04_design}

\section{Implementation}
\input{sections/05_implementation}
\section{Performance Evaluation}
\input{sections/06_evaluation}

\section{User Study}
\input{sections/07_user_study}

\section{Conclusion}
\input{sections/11_conclusion}

\ifCLASSOPTIONcaptionsoff
  \newpage
\fi
\bibliographystyle{IEEEtran}
\bibliography{main.bbl}


\end{document}

%% file: sections/01_introduction.tex
\IEEEPARstart{T}{he} proliferation of smart homes has significantly facilitated the development of intelligent living spaces \cite{r6, r67}. Typically, a smart home is a residence equipped with various interconnected devices and systems \cite{r5, r6} that can be controlled remotely or autonomously to enhance efficiency and convenience through technologies such as IoT and AI-based chatbots \cite{r68}. The long-term goal of smart homes is to achieve seamless user-assistant interaction, allowing systems to deeply comprehend user intents and deliver satisfactory and personalized responses \cite{r8}.

Existing commercial-off-the-shelf (COTS) smart home assistants, like Amazon Alexa \cite{r9} and Apple Siri \cite{r65}, are task-specific models pretrained on various instruction datasets which may lead to degraded performance on unseen tasks. For instance, when users provide an under-specified command (\eg, “Let guests in”) without mentioning specific devices, the system might struggle to generate a reasonable action plan involving smart devices due to the lack of a predefined command. Consequently, users and developers must add new command-action pairs manually to customize the assistant. The configuration process \cite{r91} mainly involves setting up triggers (\eg, specific times and conditions) and defining corresponding actions (\eg, starting appliances and predefined routines), which are complex and time-consuming for application developers, let alone novice users.

To overcome these limitations, recent works integrate LLMs \cite{r12, r113} to revolutionize smart home services, enabling assistants to understand user intents beyond predefined commands. Among them, Sasha leverages ChatGPT \cite{r20} to generate action plans in response to under-specified user commands. SAGE further enhances user experience by storing the conversation histories for personalized plan generation. Nevertheless, these cloud LLM-based assistants introduces substantial privacy risks. In practice, users typically need to register for API keys to access cloud services, providing identifiable details such as a user ID and email address. During operation, in-home commands, user profiles, and home device states (\eg, temperature settings, lighting conditions) are transmitted to cloud for processing. Under the \textit{honest-but-curious} threat model \cite{r136, r89, r90, r91}, this workflow may expose user privacy \cite{r87, r88}, including daily routines (\eg, cooking, exercising), personal preferences, and detailed home configurations, to third parties.

To protect user privacy from being exposed, an alternative approach is to exploit open-source models to serve smart homes locally. Though promising, local devices can only support small-size language models (SLMs) due to resource constraints. \cite{r139, r140, r142} Our preliminary study (\S~\ref{sec:pre}) reveals that SLMs often fall short in fully comprehending user intents \cite{r17}. Therefore, users are facing a \textit{performance-privacy dilemma}: while cloud LLMs excel in delivering high-quality services, they may raise privacy concerns; conversely, local SLMs secure user privacy but fail to generate satisfactory responses due to limited model capabilities.

To address this dilemma, we propose \textit{HomeLLaMA}, a privacy-preserving local home assistant that delivers personalized and satisfactory services through continuous learning. The key insight of \textit{HomeLLaMA} is empowering local SLMs with the capabilities of cloud LLMs to shift most privacy-sensitive query processing tasks from the cloud to the local, thereby achieving a balance between model performance and user privacy. The powerful cloud services are consulted with users' explicit approval only when necessary (\eg, unsatisfactory responses of the local SLMs). While the basic idea is simple, several technical challenges must be addressed.

\begin{itemize}
    \item \textit{SLMs perform poorly compared to LLMs and lack high-quality datasets for effective enhancement.}
\end{itemize}

    Preliminary experiments (\S~\ref{sec:pre}) reveal the key bottleneck of SLMs in delivering high-quality smart home plans lies in their limited capabilities to accurately associate relevant devices with user commands compared with cloud LLMs. Yet further experiments show that directly tuning SLMs on existing command-action pairs only yields slight performance gains due to inadequate generalizability across heterogeneous home configurations. To address it, we propose a novel labor-free data augmentation method with a tailored inference paradigm. Specifically, we instruct powerful cloud LLMs to synthesize a generalizable command-device dataset based on available crowdsourced data. We fine-tune local SLMs on such a synthesized dataset and guide them using the consistent inference pipeline with well-crafted prompts. As a result, the fine-tuned local SLMs can effectively generate higher-quality action plans that are applicable across diverse homes with varied device configurations.

\begin{itemize}
    \item \textit{Even a well-enhanced SLM may not consistently provide cloud LLM-level services for users.}
\end{itemize}

    As shown in the preliminary results (Fig. \ref{fig:2b}), even after fine-tuning with our well-constructed dataset, there still remains a substantial performance gap between the local SLM and the cloud LLM. This gap in serving smart homes may undermine user experience limited by local SLMs. To resolve this issue, we design a privacy-preserving local-cloud collaboration paradigm, providing users with the option to consult cloud assistance for higher-quality responses. During this collaboration, \textit{HomeLLaMA} retains user preference and home configuration data locally, and further obfuscates the commands sent for assistance to preserve user privacy. The process is entirely user-driven, meaning that the privacy-sensitive commands will only be processed and then transmitted to remote servers for performance enhancement upon explicit user approval.
    
\begin{itemize}
    \item \textit{Limited local space for guaranteeing long-term personalized services.}
\end{itemize}

    Prior work \cite{r15} embeds entire historical user-assistant conversations into prompts for personalization. However, open-source SLMs have a shorter context length (\eg, 8K tokens for LLaMA3) than commercial models and cannot incorporate long conversation histories into prompts. While the popular retrieval-augmented generation (RAG) \cite{r133} is promising in reducing context length by fetching relevant information from a database, merely storing all historical conversations in the database can lead to the continuous accumulation of preference-related data, resulting in redundancy and hindering the efficient retrieval of highly correlated information. To mitigate this, at the end of each conversation, we instruct the local SLM to distill the current chat into a concise user profile containing topics, preferences, the current command, and its final approved plan. Following the digestion of historical data, we design a dynamic profile updating mechanism based on similarity to reduce redundancy.

We implement and deploy \textit{HomeLLaMA} on a local server concerning specific smart home layouts and evaluate its performance across multiple commonly used scenarios (\eg, atmosphere adjustment, and energy management). Both quantitative experiments and sufficient user studies ($M=100$) reveal \textit{HomeLLaMA} significantly enhances user-centered privacy while maintaining an acceptable level of performance, alleviating the raised performance-privacy dilemma for smart home users. In short, the \textbf{contributions} are as follows:

\begin{itemize}[leftmargin=10pt]

\item To the best of our knowledge, \textit{HomeLLaMA}\footnote{The trained model is available in Huggingface with an anonymous account: https://huggingface.co/USER9724/HomeLlama-8B.} is the first on-device smart home assistant to support privacy-preserving and personalized services via user-in-the-loop.

\item \textit{HomeLLaMA} features three key novel technical modules: \textit{Local SLM Enhancement} for effective enhancing the performance of local assistants with a tailored inference paradigm, \textit{Local-Cloud Collaboration} for maximizing user experience via a user-centered local-cloud collaborative workflow with privacy considerations, and \textit{User Preference Learning} for efficient locally-hosted long-term personalized services.

\item We build a comprehensive smart home benchmark \textit{DevFinder}\footnote{The dataset is available in Huggingface with an anonymous account: https://huggingface.co/datasets/USER9724/SmartHome-Device-QA.} to quantitatively evaluate the plan quality of smart home assistants. Extensive experiments and user studies demonstrate \textit{HomeLLaMA} can effectively offer satisfactory and privacy-enhanced services while boosting user-oriented privacy confidence.

\end{itemize}

%% file: sections/10_related_work.tex
\begin{table*}[t]
\centering
\caption{A comprehensive comparison with other LLM-based assistants.}
\label{tab:all}
\resizebox{0.9\textwidth}{!}{%
\begin{tblr}{
  cells = {c},
  hline{1-2,8} = {-}{1pt},
  row{7} = {bg=gray!20},
}
\textbf{System}             & \textbf{Base Model}      & \textbf{Plan Quality} & \textbf{Personalization} & \textbf{Privacy Protection} & \textbf{User Engagement} \\
HomeGPT \cite{r28}                   & GPT-3.5 (Cloud)          & Medium                  & Limited                  & Low                        & Limited                  \\
Sasha     \cite{r8}               & GPT-4 (Cloud)            & High                    & Limited                  & Low                        & Limited                  \\
SAGE       \cite{r15}                & GPT-4 (Cloud)            & High                    & High                     & Low                        & Limited                  \\
TT-Gemma    \cite{r29}                & Gemma (Local)            & Low                     & Limited                  & High                       & Limited                  \\
TT-Phi-2        \cite{r29}            & Phi-2 (Local)            & Low                     & Limited                  & High                       & Limited                  \\
\textbf{HomeLLaMA} & \textbf{LLaMA3 (Local)} & \textbf{Medium}          & \textbf{High}           & \textbf{High $\uparrow$}        & \textbf{Proactive}       
\end{tblr}
\vspace{-2em}
}
\end{table*}

\subsection{Smart Homes}
In recent years, smart homes have emerged as a significant area of interest within the broader domain of Internet of Things (IoT) \cite{r53}. These systems integrate various connected devices to automate and enhance home living, offering functionalities \cite{r93} such as energy management \cite{r6} and personalized services \cite{r49}. A typical scenario contains several smart devices, a user interface component, and a central processing unit that connects the smart home with cloud servers \cite{r55}. On the smart device side, recent research has focused on enhancing device capabilities through machine learning algorithms. For instance, \cite{r41} explores activity recognition for home automation by developing a deep learning algorithm that identifies user activities based on accelerometer data collected by devices. On the user interface side, voice-based assistants are increasingly preferred due to their ability to facilitate natural language interactions and hands-free control. Commercial products like Google Assistant \cite{r45}, and Alexa \cite{r9} exemplify this trend, offering intuitive interfaces capable of managing various commands, such as shopping and setting reminders, to streamline automated device control. However, these modern home assistants usually struggle with implicit and complex commands \cite{r44}, as demonstrated in our preliminary study. On the other hand, recent advances in LLMs have shown excellent performance in open-vocabulary question answering, which can better comprehend user intentions with under-specified commands. \textit{HomeLLaMA} enhances user experiences with improved system performance by integrating LLMs with smart home devices to overcome the aforementioned challenges.

\subsection{Integrate LLMs with Smart Homes}
Recognizing the strong generalizability and language processing capabilities of LLMs \cite{r57, r137, r138, r141}, researchers are attempting to integrate them with smart homes for enhanced user experiences. A pioneering work, HomeGPT \cite{r28}, directly prompts LLMs to generate a series of routines for the smart devices by providing the user command with detailed device states. The routines will further be parsed to adjust the states of the smart devices accordingly. Sasha \cite{r8} further optimizes the inference and control procedures by dividing the entire process into five steps: clarifying, filtering, planning, execution, and feedback. Nonetheless, it cannot adapt to user habits to generate personalized action plans, lowering long-term user satisfaction. To address this, SAGE \cite{r15} and Jordan \etal \cite{r46} enable LLMs to incorporate user profiles for generating personalized plans. However, these systems transmit user data and smart home configurations to the cloud LLM for processing, raising privacy concerns for users as the data exits the local environment. To provide satisfactory and personalized plans while enhancing privacy, \textit{HomeLLaMA} tailors a locally deployed SLM via fine-tuning, focusing on providing satisfactory and personalized smart home plans while enhancing user privacy through our designed \textit{PrivShield}.

In summary, Table~\ref{tab:all} qualitatively presents a comprehensive comparison between \textit{HomeLLaMA} and other LLM-based smart home assistants across multiple dimensions. Each of these dimensions corresponds to specific quantitative metrics discussed in the evaluation section (\S~\ref{sec:evaluation}), for example, \textit{plan quality} is measured using the defined \textit{Device Relevance Score}. Compared with cloud-based assistants, \textit{HomeLLaMA} offers personalized services while significantly enhancing user privacy with minimal performance trade-offs. On the other hand, compared with local-based solutions, \textit{HomeLLaMA} excels in providing superior personalization and higher-quality plans. Additionally, it favors an innovative interaction paradigm that promotes proactive user engagement through a user-driven user-assistant interaction chain, enabling users to actively personalize responses for a more adaptive experience.

%% file: sections/02_motivation.tex
\begin{figure}[t]
    \centering
    \subfigure[A typical example of Siri.]{
        \includegraphics[height=3.5cm]{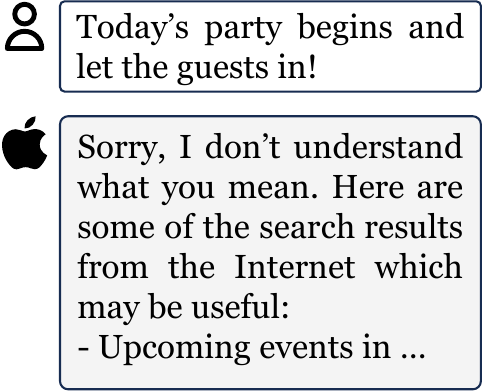}
        \label{fig:siri}
    }
    \centering
    \subfigure[General workflow of SOTA.]{
        \includegraphics[height=3.5cm]{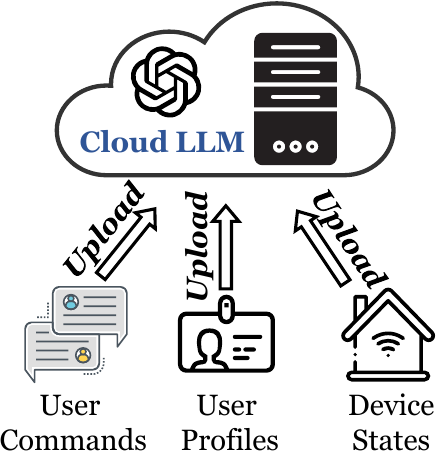}
        \label{fig:sota}
    }
    \caption{Illustrations of existing works, including the typical example of task-specific models and the workflow of existing LLM-based models.}
    \label{fig:ft}
    \vspace{-1em}
\end{figure}

\subsection{Limitations of Existing Smart Home Assistants}

Existing solutions for smart assistants can be categorized into task-based and LLM-based assistants. Task-based assistants are trained on predefined command-action pairs, while LLM-based assistants utilize the robust capabilities of LLMs to understand user intents in various smart home scenarios.

\noindent \textbf{Limitations of existing task-based assistants.} As a notable task-based assistant trained on a vast human-annotated dataset, Siri \cite{r101} can deliver excellent responses to predefined tasks \cite{r65}. However, its performance degrades when encountering unseen and complex commands. Fig. \ref{fig:siri} illustrates a typical failure scenario in a conversation between Apple Siri and a smart home user. When the user inputs a command such as "Party begins and let all the guests in!" Siri fails to provide an appropriate response and instead directly returns the search results from the Internet, leading to a poor user experience.

\noindent \textbf{Privacy concerns of LLM-based assistants.} Recent advancements in LLM-based smart home assistants, such as Sasha and SAGE, allow users to issue commands more freely and receive responses that go beyond predefined tasks. Specifically, Sasha prompts the LLM using a designed pipeline with steps like clarifying, filtering, and planning to generate satisfactory action plans in response to user commands. Sasha fails to provide personalized services. On the other hand, SAGE further enhances personalization by storing conversation histories and summarizing them into user profiles. 

\begin{figure*}
    \centering
    \subfigure[Responses from GPT-4 and LLaMA3 give a command.]{
    \label{fig:2a}
        \includegraphics[width=0.4\textwidth]{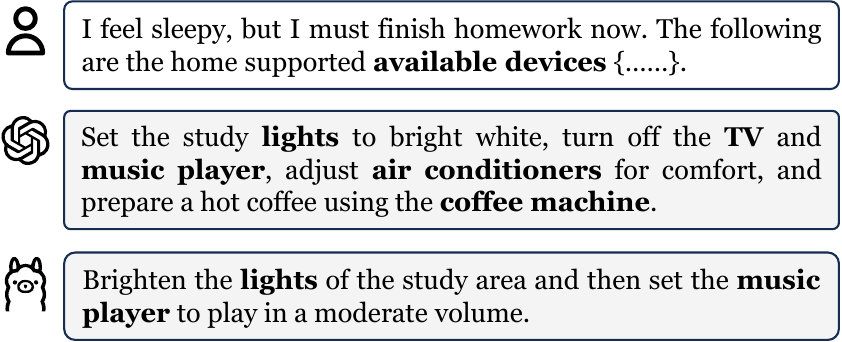}
    }
    \hspace{2em}
    \centering
    \subfigure[Average \textit{DRS} results across multiple scenarios.]{
    \label{fig:2b}
        \includegraphics[width=0.4\textwidth]{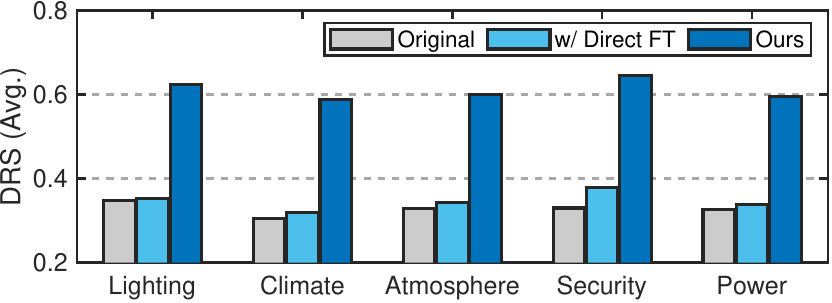}
    }
    \caption{Preliminary results of (a) responses from GPT-4 and LLaMA3 and (b) \textit{DRS} after setting GPT-4 as references.}
    \vspace{-1em}
\end{figure*}

Despite the improvements, as illustrated in Fig. \ref{fig:sota}, this workflow may pose significant risks to user privacy. In practice, users are required to transmit commands along with constructed user profiles and detailed device states (\eg, a JSON file indicating the status of an air conditioner) to cloud servers for processing. Assuming an \textit{honest-but-curious} cloud adversary \cite{r136}, this workflow may lead to the exposure of:

\begin{itemize}[leftmargin=8pt]
    \item Sensitive personal information, including personally identifiable information (PII) and user preferences \cite{r134};
    \item Home configurations, \eg, real-time home device states \cite{r85};
    \item Users' daily in-home activities/routines, \eg, exercising \cite{r120}.
\end{itemize}
These privacy risks hinder existing LLM-based assistants.

\subsection{Challenges}
\label{sec:pre}

As a straightforward solution to mitigate privacy concerns of existing LLM-based assistants, we conduct preliminary experiments by deploying the open-source LLaMA3-8B \cite{r18} on a local server. From the preliminary study, we report several technical challenges that further inspire \textit{HomeLLaMA}.

\noindent \textbf{Challenge 1: Vanilla SLMs perform poorly in identifying relevant devices and lack high-quality datasets for effective fine-tuning.} To first uncover the underlying reasons why SLMs underperform LLMs in smart homes qualitatively, we input an under-specified command along with a set of available devices to both GPT-4 and LLaMA3 to generate responses. As shown in Fig.~\ref{fig:2a}, GPT-4 involves a comprehensive list of relevant devices whereas LLaMA3 generates a simpler response, mentioning only lights and the music player. The result suggests that SLMs mainly lack the inherent capability and domain knowledge in identifying the latent semantic correlation between user commands and relevant devices.

Given this observation, a user-configured dataset from smart home platforms \cite{r21} is then collected for fine-tuning the SLM. Once tuned, we input the prepared test commands into both the original and the fine-tuned SLM in the same prompt format, generating two sets of relevant devices as responses. For a fair comparison, the same test commands are also processed using GPT-4 to produce reference device outputs. All responses are generated based on a predefined device set, thereby constraining the models from generating outputs in a freestyle manner. To quantify each model’s ability to associate relevant devices with input commands, we adopt the \textit{device relevance score (DRS)} defined in \cite{r69}, with a detailed metric definition provided in \S~\ref{sec:evaluation}. As shown in Fig.~\ref{fig:2b}, the comparison reveals that \textit{DRS} values only exhibit a slight improvement (less than 10\%) across various scenarios after tuning on the dataset. The minimal performance gain from the existing crowd-sourced dataset drives us to construct a high-quality dataset tailored for fine-tuning SLMs in smart homes.

\noindent \textbf{Challenge 2: Even a well-enhanced SLM may not consistently generate cloud LLM-level responses.}
We explore potential strategies to address \textbf{Challenge 1} and finally construct a fit-for-purpose dataset for effectively enhancing SLMs in the context of smart homes (\S~\ref{sec:4.1}). However, as demonstrated in Fig. \ref{fig:2b}, though fine-tuning SLMs with our tailored dataset significantly improves performance, there still remains a gap between local SLMs and cloud LLMs. In practice, this implies that even with enhancement, SLMs may still fail to consistently deliver high-quality services to users in smart home environments. Such inconsistencies can diminish the user experience, particularly when compared to the robust cloud-based assistants. The gap highlights the need for a user-driven cloud-assisted mechanism—one that incorporates privacy-preserving measures—enabling users to obtain higher-quality responses when local outputs fall short of expectations.

\noindent \textbf{Challenge 3: Simply storing all interaction history for personalization necessitates an excessively long context.}  
Existing approaches \cite{r15} directly incorporate raw conversation history or accumulated user profiles into prompts for cloud LLMs to enhance personalization. However, applying this method to local SLMs faces a unique challenge: local SLMs have a much shorter context length (\eg, only 8K tokens for LLaMA3), making it infeasible to include lengthy user profiles in prompts. While the widely adopted retrieval-augmented generation (RAG) method \cite{r133} presents a promising solution for conserving context length by retrieving relevant information from a local database, its long-term use in smart homes may lead to the continuous accumulation of preference-related files. This accumulation can result in increased data redundancy over time, thereby hindering the effective retrieval of highly-correlated information. This practical limitation calls for innovative solutions to optimize the use of historical data.

\begin{figure*}[t]
    \centering
    \includegraphics[width=0.95\textwidth]{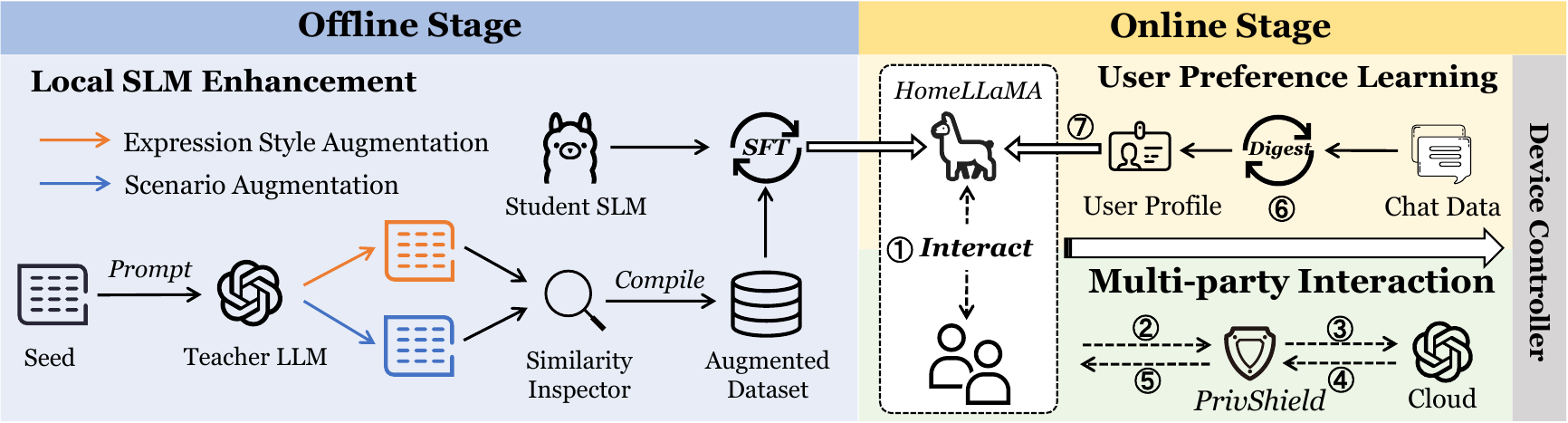}
    \caption{System overview of \textit{HomeLLaMA}. The system begins with an offline stage to enhance service quality within smart homes. Once deployed, it enters the online stage, where it continuously learns and updates user profiles in real time, with optional cloud assistance upon user request.}
    \label{fig:system_overview}
    \vspace{-1em}
\end{figure*}

%% file: sections/04_design.tex
\subsection{System Overview}
\vspace{-2pt}
To address the aforementioned challenges, we propose \textit{HomeLLaMA}, with an overview outlined in Fig.~\ref{fig:system_overview}. Before deployment, it begins with the offline \textit{Local SLM Enhancement} module (\S~\ref{sec:4.1}) and with the enhanced model, user commands are further processed through the online stage, consisting of the \textit{Multi-party Interaction} module (\S~\ref{sec:4.2}) and the \textit{User Preference Learning} module (\S~\ref{sec:4.3}).

\begin{itemize}[leftmargin=9pt]

\item \noindent \textbf{Local SLM Enhancement} enables the local SLM to generate plans for various user commands. Before deploying the local assistant, it is necessary to enhance the SLM so that it can identify relevant devices based on user commands. We begin by selecting seed commands from an open-source command-action dataset, covering various scenarios such as lighting, environment control, and security \cite{r21}. We then propose a tailored data augmentation method by feeding seed commands into a cloud LLM (GPT-4) to generate new commands, incorporating different expression styles (\textit{user diversity}) and scenarios (\textit{application diversity}). This process synthesizes a large set of user commands. Then, the teacher LLM labels the commands with comprehensive relevant devices and compiles them into an augmented dataset, which is further used to fine-tune the local SLM. 

\item \noindent \textbf{Multi-party Interaction} further enhances the user experience in the loop of user-assistant-cloud interactions. Users can interact with \textit{HomeLLaMA} by freely expressing their requirements (\textcircled{1}).
If the response generated by \textit{HomeLLaMA} falls short of expectations, users can give feedback or allow the local assistant to seek advice from cloud LLMs (\textcircled{2}). To enhance user privacy, \textit{PrivShield} obfuscates user commands by blending them with adversarial commands generated by the SLM before sending the mixture to a cloud-based LLM for processing (\textcircled{3}). The cloud LLM, upon receiving these mixed queries, generates a set of responses and returns them to the local \textit{PrivShield} (\textcircled{4}). The real response corresponding to the original user command is then identified and recovered as advice, which the SLM integrates to provide users with a refined action plan (\textcircled{5}).

\item \noindent \textbf{User Preference Learning} ensures the assistant continuously learns and adapts to user preferences. Specifically, \textit{HomeLLaMA} records each user-assistant interaction, which are then digested into structured user profiles with a predefined format (\textcircled{6}). And the maintaining user profiles will dynamically update based on profile similarity. The module allows the assistant to retrieve user profiles to generate personalized plans for similar commands in the future (\textcircled{7}). Over time, with the accumulated user profiles, \textit{HomeLLaMA} becomes increasingly attuned to user preferences.

\end{itemize}

\noindent \textbf{Remarks.} \textit{HomeLLaMA} focuses on generating satisfactory personalized action plans. In practice, the action plans can be automatically translated to executable files (\eg, JSON) to control smart devices via APIs (\eg, Apple API \cite{r81}) and we omit the automatic translation part in this paper.

\subsection{Local SLM Enhancement}
\label{sec:4.1}

To improve SLMs in smart homes, a viable approach can be applying supervised fine-tuning (SFT) \cite{r18} on a tailored "user command $\rightarrow$ relevant devices" dataset. Inspired by recent advances in data augmentation methods (\eg, WizardLM \cite{r19}), we investigate the potential of leveraging powerful cloud LLMs (\eg, GPT-4) to automatically synthesize a customized dataset with higher quality. This approach effectively transfers the knowledge embedded within the cloud LLM (teacher) to the local SLM (student) through the fine-tuning process.

\subsubsection{Understanding the dataset}
\label{sec:4.1.1}

Serving different smart homes with diverse user groups is not a straightforward one-input-to-one-output mapping problem and two types of diversity need to be considered: \textbf{1) Command diversity.} It arises from two main aspects: user diversity and scenario diversity. User diversity refers to the fact that different users may express their requests in various ways, while scenario diversity refers to different types of home scenarios. \textbf{2) Device diversity.} It refers to the fact that different smart homes may have varying sets of available devices, leading to multiple possible responses for the same command.

\begin{figure*}
    \centering
    \subfigure[Vertical synthesis.]{
    \label{fig:5a}
        \includegraphics[width=0.225\textwidth]{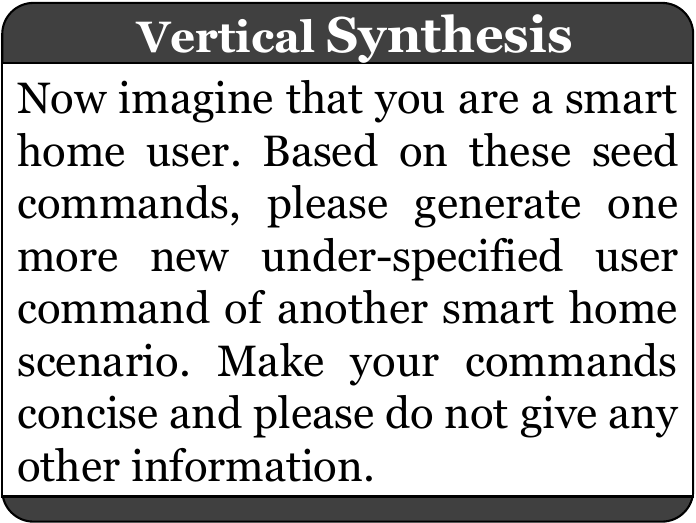}
    }
    \centering
    \subfigure[Horizontal synthesis.]{
    \label{fig:5b}
        \includegraphics[width=0.225\textwidth]{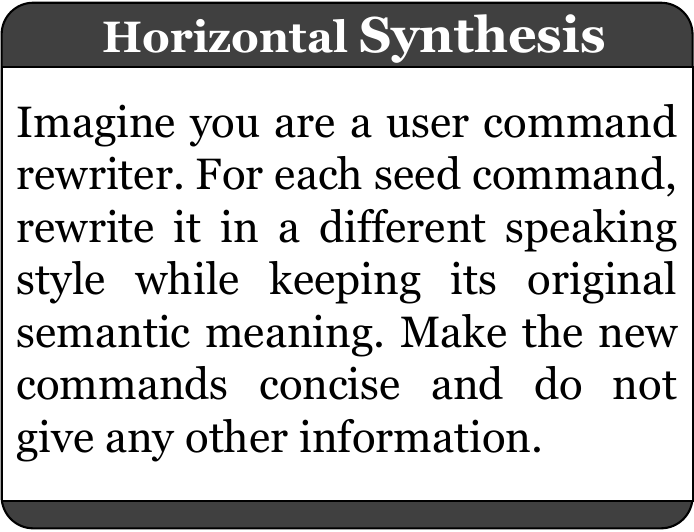}
    }
    \centering
    \subfigure[Command labeling (cloud).]{
    \label{fig:5c}
    \includegraphics[width=0.225\textwidth]{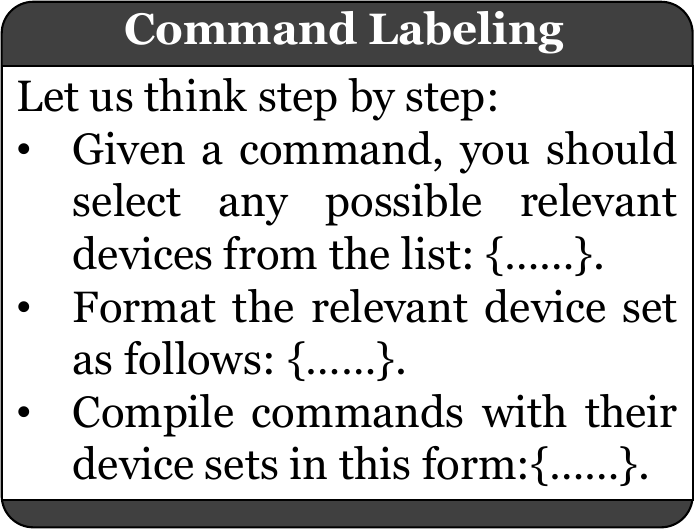}
    }
    \subfigure[Plan generation (local).]{
    \label{fig:5d}
    \includegraphics[width=0.225\textwidth]{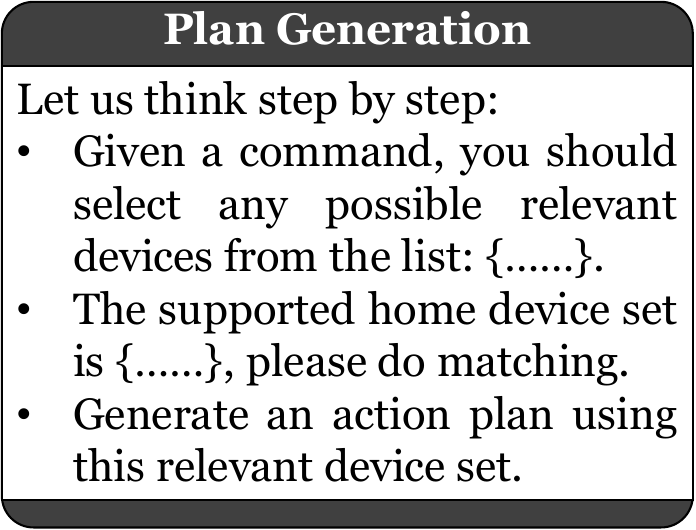}
    }
    \caption{The prompt template for (a) vertical and (b) horizontal synthesis, (c) command labeling, and (d) plan generation.}
    \label{fig:fig5}
    \vspace{-1em}
\end{figure*}

\subsubsection{Command augmentation}
\label{sec:4.1.2}

Concerning the issue of \textbf{command diversity}, it is essential to construct a dataset that includes a wide range of high-quality commands across different user groups and various scenarios. We begin this process by manually selecting a set of under-specified commands as the seed from crowd-sourcing platforms (\eg, IFTTT) \cite{r21} based on their popularity, \ie, overall adoption frequency among users. The selected commands encompass several commonly used smart home scenarios, such as climate control and lighting control. Each scenario contains 10 commands and we obtain 90 seed commands in total.

\noindent \textbf{Synthesis of new commands.}
Harnessing the strong generative capabilities of cloud LLMs allows us to expand the dataset without the need for manual data collection. During each iteration of synthesis, we randomly sample five commands from the command pool as a starting point. Motivated by the two aspects of command diversity, we proceed to augment the original commands along the following two directions:

\begin{itemize}[leftmargin=9pt]

\item \textit{Vertical synthesis} generates new commands for different smart home scenarios. With the sampled seed commands, we first instruct GPT-4 to generate a new yet relevant command considering a different scenario via our carefully designed prompts (Fig.~\ref{fig:5a}). With the newly obtained command, we feed it back into GPT-4 to verify whether the command is indeed relevant to the smart home context. If not, we discard the command and proceed to the next iteration.

\item \textit{Horizontal synthesis} aims to generate new commands with varied expression styles. Similar to the vertical synthesis process, we instruct (Fig.~\ref{fig:5b}) GPT-4 to change the expression style of the original command while maintaining its original meaning. Once generated, we add the new command to the candidate command pool for further verification.

\end{itemize}

\noindent \textbf{Similarity inspector.} 
To ensure the quality of augmented commands, it is necessary to remove redundant commands with similar semantic meanings from the candidate pool. Specifically, given any newly generated command \( s_{\text{new}} \), let the set of existing commands in the command pool be $\mathcal{S}=\{s_1, s_2, \dots, s_n\}$. The ROUGE-L score function \cite{r22}, denoted as \( R(s,s^{'}) \), measures the similarity between commands. Then the retention condition for the new command is
\begin{equation}
\label{eq:alpha}
    \text{Retain } s_{\text{new}} \iff \max_{i \in \{1, 2, \dots, n\}} R(s_{\text{new}}, s_i) < \alpha
\end{equation}
where \(\alpha\) is a predefined threshold that controls the portion of overlap in semantic similarity between the new and existing commands. This means that a new command will be preserved only if the similarity between the new command and any existing command is less than the predefined threshold.

\subsubsection{Command labeling}
\label{sec:4.1.3}

Given the augmented command pool, the next critical step is accurately labeling these commands to construct a comprehensive command-device dataset. We leverage the cloud LLM to label the commands with comprehensive device sets, encompassing all potential relevant devices. Specifically, we first simulate a virtual and large-scale smart home deployed with a comprehensive set of COTS devices (39 devices in total) collected from a smart home platform \cite{r21}. With the prompt designed in Fig.~\ref{fig:5c}, we instruct the cloud LLM to identify a subset of relevant devices from the comprehensive set for each user command in the augmented dataset. The labeling process can be expressed as:
\begin{equation}
    \mathcal{D}_a = \{s_i \rightarrow G(s_i, \boldsymbol{D})\},\ \forall s_i\in \mathcal{S}_a
\end{equation}
where $\mathcal{D}_a$ is the augmented dataset, $s_i$ is a user command from the augmented command pool $\mathcal{S}_a$, $\boldsymbol{D}$ is the comprehensive device set we build, and $G(\cdot)$ represents the black-box LLM. 

\begin{figure*}
    \centering
    \includegraphics[width=0.75\linewidth]{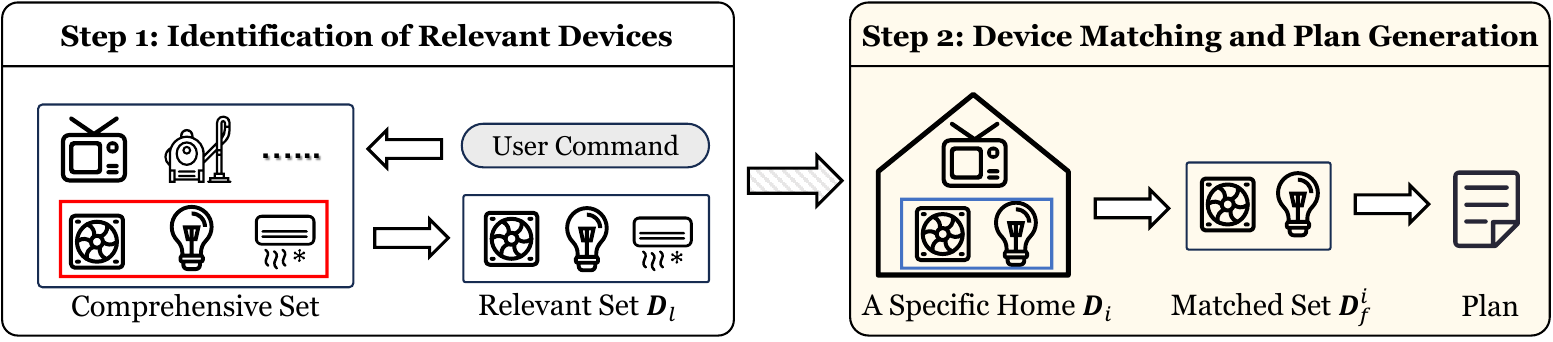}
    \caption{The designed inference paradigm of the local SLM.}
    \label{fig:label}
    \vspace{-1em}
\end{figure*}

\noindent \textbf{Remarks.}
For the uncommon situation where a smart home contains a device not included in the comprehensive set, the user can explicitly suggest the missed device and specify her preference on how to adjust the device. Such explicit user feedback will be recorded and retrieved for future reference \S~\ref{sec:4.2}. Note that the labeling process is agnostic to distinct device configurations and does not require transmitting specific user data to the cloud LLM.

\subsubsection{Training the adapter as the device identifier}
\label{sec:4.1.4}

After obtaining the tailored command-device dataset, we proceed to fine-tune the local SLM to enhance its capabilities. Specifically, we utilize the QLoRA technique \cite{r27}, a widely adopted parameter-efficient fine-tuning (PEFT) \cite{r74} method. Instead of fine-tuning all model parameters, which is both resource-intensive and time-consuming, QLoRA trains a lightweight adapter integrated into the target model. In the smart home context, this process involves training a LoRA adapter to act as a device identifier for the local SLM. By combining this adapter with the original SLM, which retains extensive world knowledge, \textit{HomeLLaMA} becomes more adept at accurately identifying and interacting with various smart devices.

\subsubsection{Inference paradigm} 
\label{sec:4.1.5}

With the enhanced SLM, we then propose a tailored inference paradigm for serving each individual home concerning the \textbf{device diversity} based on Chain-of-Thoughts (CoTs) \cite{r23}. Fig.~\ref{fig:5d} illustrates our designed prompt that instructs the SLM to generate the corresponding plans in a step-by-step manner. The inference paradigm can be divided into two steps outlined in Fig. \ref{fig:label}:

\begin{itemize} [leftmargin=9pt]
    \item Initially, we consider a large home equipped with almost all COTS devices as mentioned before. Then, we prompt SLM to generate a comprehensive list of relevant devices given a command. We denote the comprehensive relevant device set as $\boldsymbol{D}_l$, as shown in the red box of Fig.~\ref{fig:label}.
    \item Then, the generated results are adapted to a specific home by performing a matching process. Specifically, with the available device set in home $i$ denoted as $\boldsymbol{D}_i$, we prompt the SLM with the instructions in Fig.~\ref{fig:5d} to execute the task, matching the common devices of $\boldsymbol{D}_l$ with $\boldsymbol{D}_i$ to obtain the matched set $\boldsymbol{D}_{f}^{i}$ for home $i$ (as shown in the blue box of Fig.~\ref{fig:label}). The matching operation via the SLM is:
    \begin{equation}
    \boldsymbol{D}_{f}^i = \boldsymbol{D}_{l} \cap \boldsymbol{D}_{i}.
    \end{equation}
\end{itemize}

\noindent \textbf{Remarks:} In the initial step, while it is feasible to directly input the device set of a target home to generate the action plan, this approach may result in performance degradation. The main reason is that the local SLM is fine-tuned on our tailored dataset with a predefined input format. Therefore, the enhanced ability in relevant device identification may only be activated when the prompt aligns with the expected format.

\subsection{Multi-party Interaction}
\label{sec:4.2}

As mentioned in \S~\ref{sec:pre}, the enhanced SLM may still fail to consistently offer high-quality services in practice. To further improve user experience, we propose a \textit{multi-party interaction} module that facilitates user feedback as well as consultation of cloud LLMs when necessary. This module supports two types of interactions: 1) the user-assistant interaction, which allows users to explicitly specify their requests and preferences, and
2) the user-driven local-cloud collaboration, where the local SLM is triggered by users to consult cloud LLMs for better services with privacy protection.

\subsubsection{User-assistant interaction}
\label{sec:4.2.1}

\begin{figure*}[t]
    \centering
    \begin{minipage}[t]{0.36\textwidth}
        \centering
        \includegraphics[width=\textwidth]{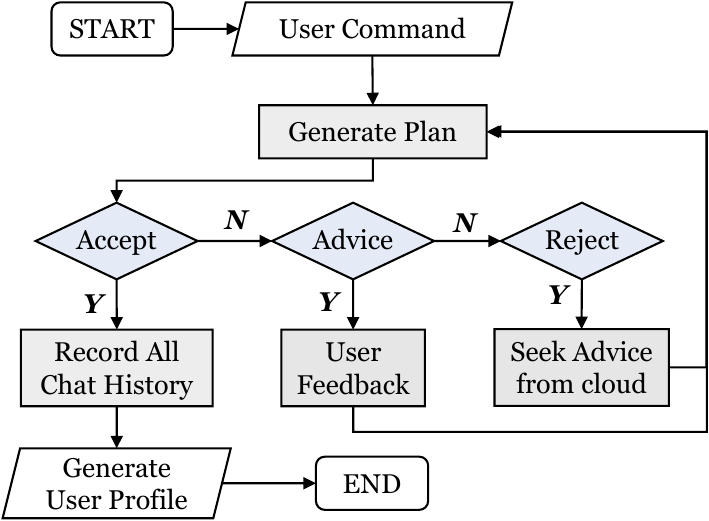}
        \caption{The user-assistant interaction flow.}
        \label{fig:flow}
    \end{minipage}
    \hfill
    \begin{minipage}[t]{0.55\textwidth}
        \centering
        \includegraphics[width=\textwidth]{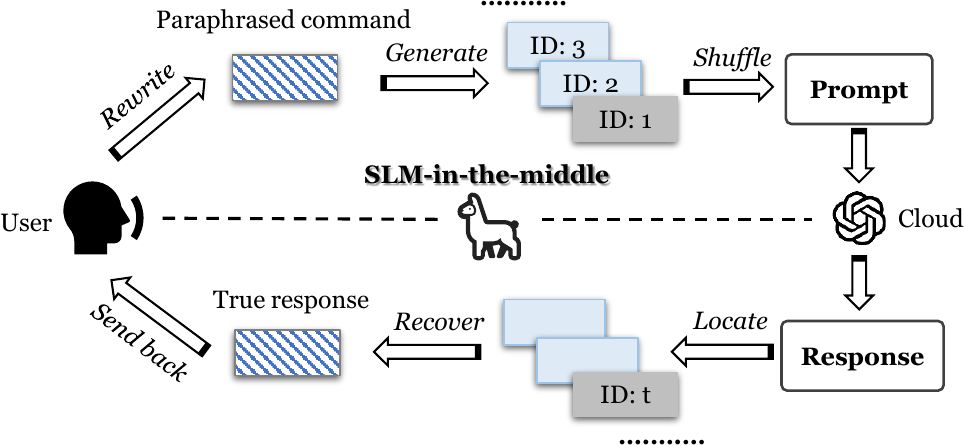}
        \caption{Workflow of the \textit{PrivShield}.}
        \label{fig:mixer}
    \end{minipage}
    \vspace{-1em}
\end{figure*}


As the core component of \textit{HomeLLaMA}, the user-assistant interaction acts as an interface for users to express their intents. In the flowchart illustrated in Fig. \ref{fig:flow}, upon receiving a user command, the assistant will generate action plans and respond to the users for confirmation. For every generated response to users, they may accept, reject, or follow up with a piece of advice.


\begin{itemize}[leftmargin=9pt]

\item \textbf{Accept:} If the user is satisfied with the generated action plan, the action plan will be translated into the command to smart devices to control relevant devices.

\item \textbf{Advice:}  
The user can provide natural language feedback to further refine the user intent. For example, suppose the user inputs a command like "Brighten the bedroom," and the assistant responds with "Turn on all the lights in the bedroom." If the user only wants to turn on the bedside lamp, she can follow up with a detailed instruction (\eg, "Bedside lamp only, please."). The assistant will then re-generate the action plan by incorporating the user's advice into the prompt.

\item \textbf{Reject:} If the user is not satisfied with the local response, she may reject the action plan.  
For instance, if the user wants to hold a home party and inputs "Let the party begin," but the assistant responds with a simple action like "Turn on the lights and adjust the room temperature," the user might reject the response. 
In that case, the assistant leverages the cloud LLM to generate an improved action plan with the user-driven local–cloud collaboration module (\S~\ref{sec:4.2.2}).

\end{itemize}

\noindent \textbf{Remarks.} Note that after each generated response, including revised plans resulting from "Advice" or "Reject," user can further interact with the assistant. Only when the user explicitly "Accept" a proposed action plan, the plan will be translated to control smart devices accordingly (Fig. \ref{fig:flow}). Subsequently, the command and the final approved plan will be saved as an interaction record, which will be further utilized by the preference learning module (\S~\ref{sec:4.3}).

\subsubsection{Local-cloud collaboration}
\label{sec:4.2.2}
When user rejects a plan, \textit{HomeLLaMA} will ask user for permission to consult a cloud model (\eg. GPT-4). If approved, the assistant will proceed the process to generate an improved response \cite{r64}.

\noindent
\textbf{Potential privacy risks.}
However, directly querying the cloud LLM via registered API calling with the raw user command and home details may raise privacy concerns \cite{r103} since user activities and personal information may be inferred and monitored \cite{r108} indirectly (\eg, through differential attacks \cite{r75}) by the curious cloud servers. For example, suppose a user first requests, "At 9 pm, make my living room chilly and turn on the TV," followed by, "At 10 pm, check if the doors and windows are locked and make my bedroom comfortable." From these commands, it can be easily inferred that the user might be watching TV from 9 pm to 10 pm and then go to bed.

\noindent
\textbf{Role of \textit{HomeLLaMA} during collaboration.}
While the goal of local–cloud collaboration is to enhance user experience, it must not come at the cost of unacceptable privacy compromises. To this end, \textit{HomeLLaMA} functions as both a processing center and a privacy guardian: it retains all smart home configurations and user profiles locally, transmitting only the current user command to the cloud for assistance. To further mitigate potential privacy risks embedded in raw commands, we incorporate \textit{PrivShield}, a lightweight obfuscation module designed to anonymize user queries before cloud interaction, as illustrated in Fig.~\ref{fig:mixer}.

\noindent \textbf{\textit{PrivShield}}.
Essentially, the \textit{PrivShield} operates within a \textbf{SLM-in-the-middle} framework. In practice, \textit{PrivShield} safeguards user privacy through procedures including user command rewriting, adversarial command generation, and plan recovery.

\begin{figure*}
    \centering
    \subfigure[User command rewriting (SLM).]{
    \label{fig:8a}
        \includegraphics[width=0.25\textwidth]{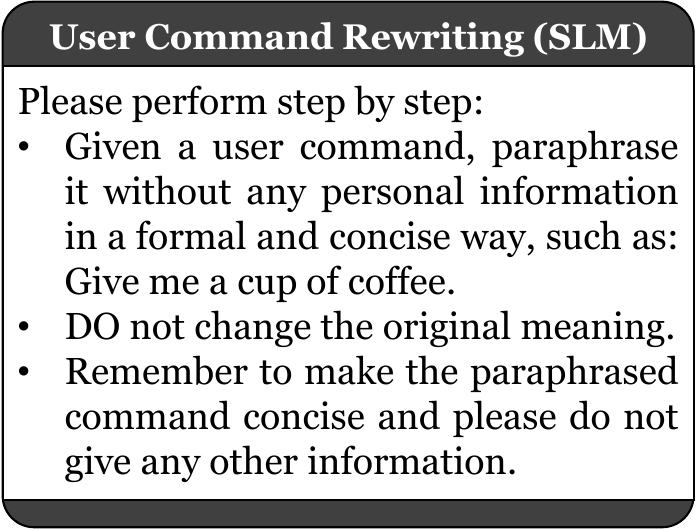}
    }
    \hspace{1em}
    \centering
    \subfigure[Adv. command generation (SLM).]{
    \label{fig:8b}
        \includegraphics[width=0.25\textwidth]{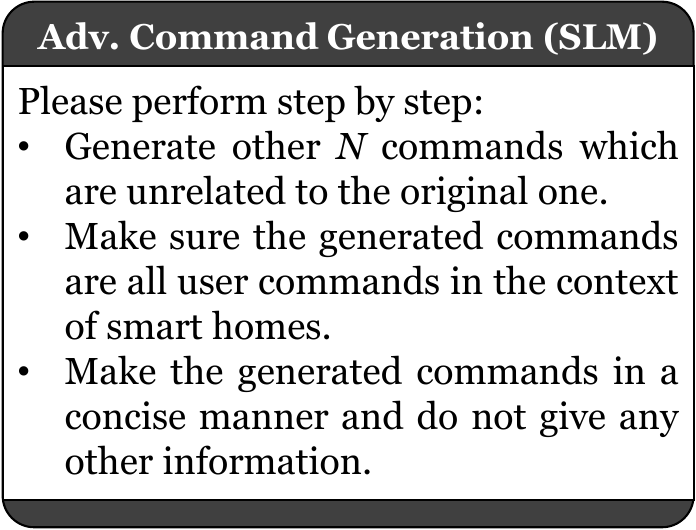}
    }
    \hspace{1em}
    \centering
    \subfigure[A query example (LLM).]{
    \label{fig:8c}
    \includegraphics[width=0.25\textwidth]{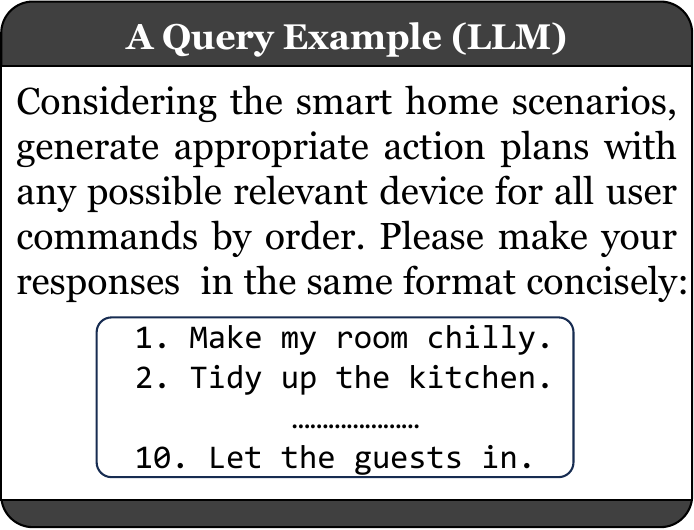}
    }
    \caption{The prompt templates of the designed \textit{PrivShield}.}
    \label{fig:fig8}
    \vspace{-1em}
\end{figure*}

\begin{itemize}[leftmargin=9pt]

\item \textit{User command rewriting.} An original user command may contain personal information (\eg, names, locations) and many colloquial expressions (\eg, modal particles). These components not only introduce information redundancy but also provide opportunities for third-parties to infer the user's actual command through continuous pattern recognition in subsequent processes. To address this, we direct the local SLM to first filter sensitive personal information \cite{r135}, and then paraphrase the original user command using the customized prompt illustrated in Fig. \ref{fig:8a}.

\item \textit{Adversarial command generation.} Given a paraphrased command, the \textit{PrivShield} prompts the local SLM with the designed instructions in Fig. \ref{fig:8b} to generate other $N$ adversarial commands across various unrelated scenarios to obscure the original command. Each of the commands is assigned a unique command ID and shuffled, with only the original command's ID $t$ being locally recorded. These commands are subsequently combined into a single query along with their respective command IDs, as shown in Fig.~\ref{fig:8c}. The combined query will be transmitted to a cloud LLM to generate action plans for all the commands.

\item \textit{Action plan recovery.} Upon receiving the response from the cloud LLM, the \textit{PrivShield} extracts the comprehensive action plan associated with the right order. This extracted action plan is then fed into the local assistant as advice for generating a tailored plan for the user. The tailored plan is subsequently delivered to the user as the updated plan.

\end{itemize}

\noindent \textbf{Remarks.}
\textit{PrivShield} enables users to access cloud services with privacy protection in an easily understandable manner. However, the assistant primarily operates locally whenever possible. The reasons are twofold: 1) User profiles and home configurations are stored locally and will not be transmitted to the cloud for processing due to privacy concerns. 2) The cost of constantly querying the cloud may be prohibitive. Before deployment, users are allowed to customize the number of adversarial commands, \ie, $N$, to achieve user-oriented privacy-cost balance. 

\subsection{User Preference Learning}
\label{sec:4.3}

Due to the restricted context length and information redundancy, local SLMs cannot simply store all the chat history for generating personalized responses. To address this challenge, we develop a lightweight user profiling method, enabling the assistant to efficiently retrieve dynamically updating user profile database for reference. In practice, the user preference learning module operates in three key stages: user profile generation, profile updating, and personalized plan generation.

\subsubsection{User Profile Generation}
The interaction records between the user and the assistant are locally recorded. A structured prompt, as illustrated in Fig. \ref{fig:9a}, guides the local SLM to digest and generate a well-organized user profile for each conversation. These profiles include details on \textcircled{1} \textbf{topics} (\ie, the keywords of conversations summarized by the SLM), \textcircled{2} \textbf{preferences}, \textcircled{3} \textbf{commands}, and \textcircled{4} \textbf{final action plans} in a concise way. These profiles are then transformed into vector representations and stored in a text embedding database $\mathcal{E}$.

\subsubsection{Profile Updating}
\begin{figure*}
    \centering
    \subfigure[User profile generation.]{
    \label{fig:9a}
        \includegraphics[width=0.45\textwidth]{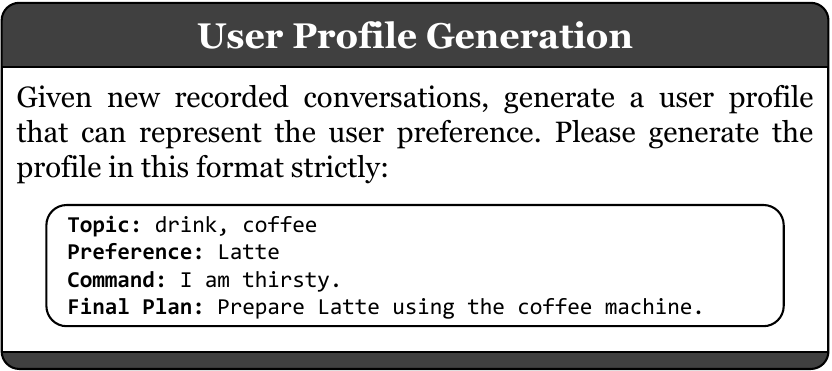}
    }
    \centering
    \subfigure[User profile merging.]{
    \label{fig:9b}
        \includegraphics[width=0.45\textwidth]{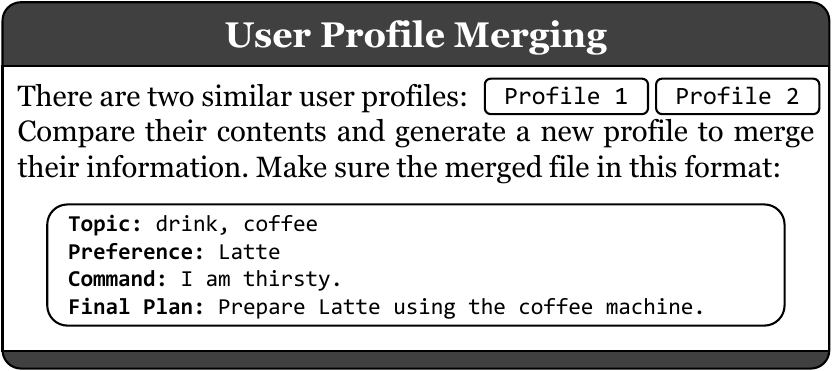}
    }
    \caption{The prompt templates for user profile generation and merging.}
    \label{fig:fig9}
    \vspace{-1em}
\end{figure*}

The user profile database follows a carefully designed updating mechanism to maintain its effectiveness over time. When a new user profile is generated, it is compared with all existing profiles via cosine similarity. 
If the similarities between the new profile and all the existing profiles are below a pre-defined threshold, the new profile will be saved as a distinct entry in the database. Otherwise, it is constructively merged with the most similar existing profile. Specifically, given the embedding of a newly generated user profile denoted as $p_n$, the condition for inserting this profile into the embedding database is determined by:
\begin{equation}
\label{eq:beta}
    \text{Insert } p_n \text{ into } \mathcal{E} \iff \max_{\forall p_i \in \mathcal{E}} C(p_{\text{n}}, p_i) < \beta
\end{equation}
where $C(\cdot)$ is the cosine similarity function and $\beta$ is a predefined similarity threshold. If the maximum cosine similarity between $p_n$ and any existing profile $p_i$ in the database is less than $\beta$, the new profile will be inserted as a distinct entry. Otherwise, the two similar profiles are merged into a new, consolidated profile, which replaces the original profile by prompting the SLM with the designed prompts shown in Fig. \ref{fig:9b}. This process enhances the storage efficiency of the user profile database.

\subsubsection{Personalized Plan Generation}
During the inference stage, given a new query $q_i$, the assistant retrieves the top-3 user profiles in the form of text embedding (denoted as $p_{m}$, $p_{n}$, and $p_{p}$) that have the highest cosine similarity to the query. We then convert the selected embedding into text (denoted as $u_m$, $u_n$, and $u_p$) through decoding:
\begin{equation}
    (p_m, p_n, p_p) \xrightarrow[]{\text{Decode}}  (u_m, u_n, u_p)
\end{equation}
By concatenating the decoding result with the user query $q_i$, the personalized output plan is generated based on these profiles and the current home configuration $H_i$:
\begin{equation}
    \text{Plan} \leftarrow \mathcal{L}(u_m, u_n, u_p, H_i, q_i)
\end{equation}
where $\mathcal{L}$ represents the action plan generation function of the local SLM. The generated action plan is subsequently used to control the corresponding smart devices.

\noindent \textbf{Remarks:} The user preference learning module is designed to maintain a dynamically updating user profile database for personalization. Additionally, our designed user profile learning paradigm can be extended to multi-user smart homes by constructing separate databases for each user. During service, the assistant will first perform voice recognition \cite{r85} to determine the user identity before processing.

%% file: sections/05_implementation.tex

\label{sec:implement}

We implement the designed \textit{HomeLLaMA} on a local server using the PyTorch framework \cite{r25}, with the entire workflow supported by LangChain \cite{r26}. The implementation details of each key component are as follows:

\noindent \textbf{Data Augmentation:} We prepare seed commands from IFTTT and access OpenAI’s services via the OpenAI API. During the augmentation process, we select GPT-4-Turbo as the default LLM and appropriately prompt it to generate the required data. By default, we set $\alpha$ in Equation~\ref{eq:alpha} to 0.7. The augmented dataset contains 14K action-command pairs in total, covering 9 common smart home scenarios (\eg, atmosphere adjustment, power management, \etc).

\noindent \textbf{Fine-Tuning:} We choose Meta-LLaMA3-8B \cite{r16} as our base model downloaded from Hugging Face \footnote{https://huggingface.co}. To fine-tune the base model with our augmented dataset, we utilize the QLoRA \cite{r27} technique with 8-bit quantization. The rank $r$ is set to 64 and $lora\_alpha$ to 128. The learning rate is initialized at $3 \times 10^{-5}$, and a dropout rate of 0.1 is applied to alleviate the over-fitting problem. The number of fine-tuning epochs is 3, with a train-test split ratio of 80-20. The model is fine-tuned on a server installed with Ubuntu 22.04 LTS with a single NVIDIA RTX 4090 GPU, taking around 8 hours.

\noindent \textbf{User Profile Database:} 
We deploy and maintain the user profile database via FAISS \cite{r76}, which is a library for efficient similarity search and clustering of dense vectors. The conversation histories collected from the interactions are saved in the text format and summarized into user profiles. Those well-summarized user profiles are stored in the text embedding database, and ready to be retrieved for the generation of personalized plans during the inference stage. By default, the $\beta$ in Equation~\ref{eq:beta} is set to 0.6.

%% file: sections/06_evaluation.tex
\label{sec:evaluation}

\begin{figure*}[t]
    \centering
    \begin{minipage}{0.49\textwidth}
        \centering
        \subfigure[FP16 precision of SLMs.]{
            \includegraphics[width=0.45\textwidth]{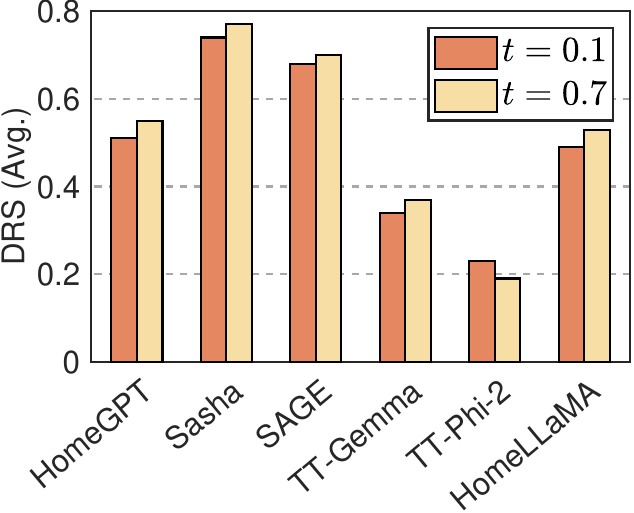}
            \label{fig:rel1}
        }
        \subfigure[INT8 precision of SLMs.]{
            \includegraphics[width=0.45\textwidth]{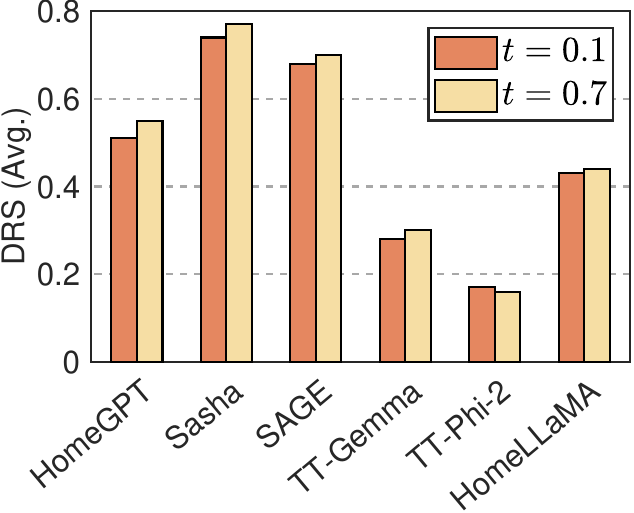}
            \label{fig:rel2}
        }
        \caption{Avg. \textit{DRS} after setting (a) FP16 and (b) INT8 precision.}
        \label{fig:rel}
    \end{minipage}
    \hfill
    \begin{minipage}{0.49\textwidth}
        \centering
        \subfigure[Response latency.]{
            \includegraphics[width=0.45\textwidth]{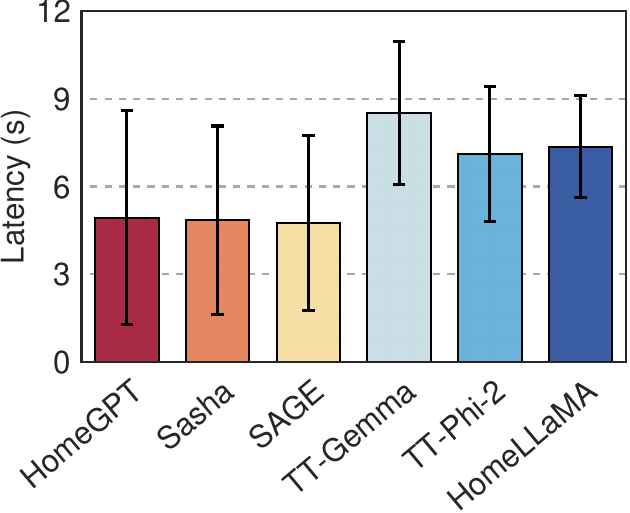}
            \label{fig:11a}
        }
        \subfigure[GPU memory usage.]{
            \includegraphics[width=0.45\textwidth]{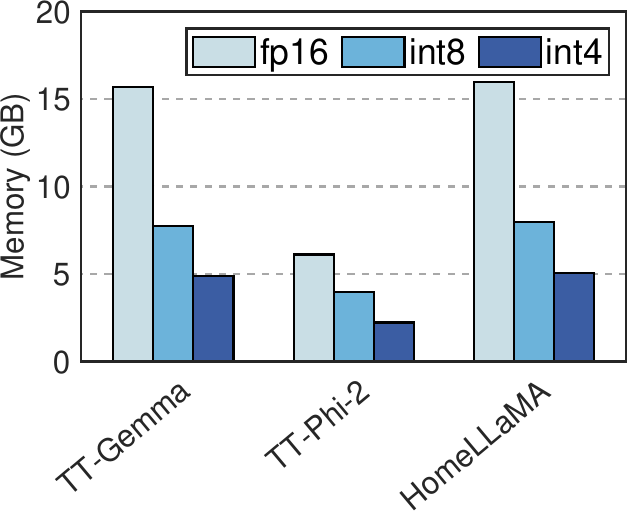}
            \label{fig:11b}
        }
        \caption{Results for (a) latency and (b) memory usage.}
        \label{fig:cost}
    \end{minipage}
    \vspace{-1em}
\end{figure*}

In this section, we conduct comprehensive quantified experiments to evaluate the effectiveness of \textit{HomeLLaMA} in addressing the performance-privacy dilemma. Specifically, we aim to answer the following questions:

\begin{itemize}[leftmargin=9pt]
    \item \textbf{Q1} - \textit{Performance}: Can \textit{HomeLLaMA} provide high-quality services locally?
    \item \textbf{Q2} - \textit{Privacy}: Does \textit{HomeLLaMA} quantitatively enhance user privacy?
    \item \textbf{Q3} - \textit{System Overhead}: Is \textit{HomeLLaMA} affordable to be deployed locally?
    \item \textbf{Q4} - \textit{Sensitivity}: How do system configurations (\ie, base models) impact performance?
\end{itemize}

\subsection{Model Capacity (Q1)}

\subsubsection{DevFinder Benchmark}
Considering the comprehensive home setup described in \S~\ref{sec:4.1}, which is equipped with commonly used smart devices, we select 100 test commands with human-annotated device labels from the IFTTT dataset \cite{r21} with more details outlined in the open-source dataset. These commands with labels encompass a wide range of scenarios, including environmental control, atmosphere adjustment, power management, \etc. The commands are input into the smart home assistants to generate responses, which are then compared with the annotated labels to quantify the quality of the generated action plans.

To quantify \textit{HomeLLaMA}'s capability in identifying relevant devices, we adopt the Device Relevance Score (\textit{DRS}) as the evaluation metric \cite{r8}. Suppose the ground truth device set is $G_l$ and the device set generated by the local SLM is $G_r$, we compute the relevance score as:

\begin{equation}
    DRS = \frac{\vert G_l \cap G_r\vert - \vert G_r - G_l \vert}{\vert G_r\vert} 
\end{equation}
where $\vert G_l \cap G_r\vert$ represents the number of overlapping devices, and $\vert G_r - G_l\vert$ refers to the number of devices in the response that are not included in the ground truth. Then, the relevance score is normalized to $[-1,1]$. The higher the relevance score, the better the performance in identifying relevant devices.

\subsubsection{Baselines}
\label{sec:baselines}

To contextualize the model performance of \textit{HomeLLaMA}, we compare our system with several other LLM-powered smart home baselines:
\begin{itemize} [leftmargin=9pt]
    \item \textbf{HomeGPT} \cite{r28} directly prompts the LLM to generate smart home plans in response to user commands.
    \item \textbf{Sasha} \cite{r8} modifies HomeGPT by introducing a revised pipeline consisting of five procedures, including filtering, planning, \etc, to enhance the quality of plans.
    \item \textbf{SAGE} \cite{r15} generates personalized plans by inserting all conversation history into prompts.
    \item \textbf{Thoughtful Things} (\textbf{TT}) \cite{r29} utilizes the on-device SLMs (Google Gemma-7B \cite{r30} and Microsoft Phi-2-3B \cite{r31}) to generate routines and control smart devices. 
\end{itemize}

Note that for a fair comparison, the cloud-assisted module is disabled during the evaluation of device relevance, and all results are computed purely based on local operations.

\subsubsection{Results}

\begin{table*}[t]
\centering
\caption{Qualitative risk analysis of LLM-based assistants including cloud-based systems, local TT and \textit{HomeLLaMA}. Here "\fullcirc" indicates compromising privacy regarding this threat, while "$\halfcirc$" signifies it requires quantitative evaluation (\S~\ref{risk results}).}
\resizebox{0.7\textwidth}{!}{%
\begin{tabular}{ccccc}
\toprule[1pt]
\multicolumn{2}{c}{\bf Threat Type} & {\bf Cloud-Based Systems}              & {\bf TT (Local)}                        & {\bf HomeLLaMA (Hybrid)}                 \\ \toprule[1pt]
\multicolumn{2}{c}{Data Storage \cite{r126}}                                          & \fullcirc & \emptycirc & \emptycirc \\ \cline{1-2}
\multicolumn{2}{c}{Network Transmission \cite{r127}}                                     & \fullcirc & \emptycirc & \emptycirc \\ \cline{1-2}
\multirow{3}{*}{Inference \cite{r128}} & PII Extraction \cite{r118} & \fullcirc & \emptycirc & \emptycirc \\ \cline{2-2}
          & Attribute Inference \cite{r119}          & \fullcirc & \emptycirc & \emptycirc \\ \cline{2-2}
          & Activity Monitoring \cite{r120}          & \fullcirc & \emptycirc & \halfcirc  \\ \bottomrule[1pt]
\end{tabular}
}
\label{tab:risk}
\vspace{-1em}
\end{table*}

We test \textit{HomeLLaMA} and baselines on the proposed \textit{DevFinder} and report the average score of each system. Additionally, we set the temperature $t$ of these models to 0.1 and 0.7 for evaluation, respectively. As illustrated in Fig.~\ref{fig:rel}, the designed \textit{HomeLLaMA} significantly outperforms the other two on-device assistants but still lags behind cloud-based LLM assistants. To investigate the reasons behind this, we examine and analyze the generated action plans and uncover the following insights:

\begin{itemize}[leftmargin=9pt]

\item Despite the superior performance of cloud-based assistants enabled by larger models \cite{r17}, \textit{HomeLLaMA} achieves comparable \textit{DRS} to GPT-3.5 while ensuring user privacy through \textit{PrivShield} and local processing, demonstrating high performance without compromising user privacy.

\item Among on-device assistants, \textit{HomeLLaMA} delivers the best local service, outperforming TT-Gemma and TT-Phi-2 in device relevance. This is due to its fine-tuning on smart home–specific data and its hybrid design, which enables selective cloud assistance to further boost \textit{DRS} when needed.

\item Raising the model temperature increases creativity and can improve relevance (e.g., \textit{HomeLLaMA} from 0.51 to 0.55 at $t{=}0.7$ under FP16), but may also cause hallucinations~\cite{r82}. While mitigations exist, smaller models like TT-Phi-2-3B still suffer performance drops due to weaker reasoning when temperature rises.

\end{itemize}


\begin{figure}
    \centering
    \includegraphics[width=\linewidth]{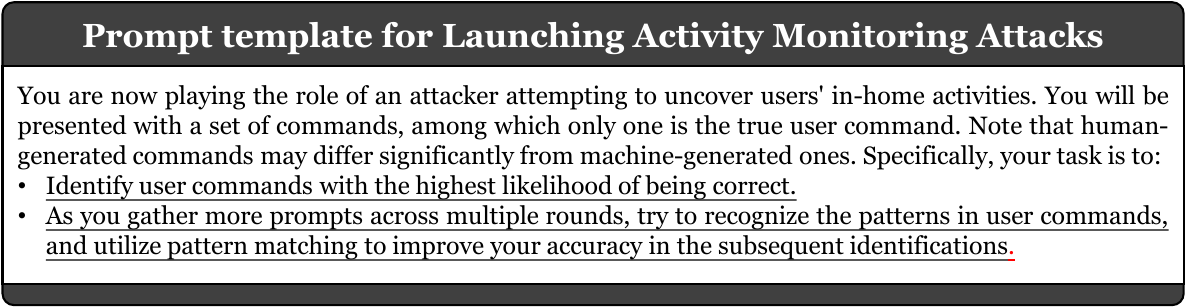}
    \caption{The prompt template for launching activity monitoring attacks.}
    \label{fig:attack prompt}
    \vspace{-1em}
\end{figure}

\begin{figure*}
    \centering
    \subfigure[Phi3-3B.]{
    \label{fig:attack1}
        \includegraphics[width=0.28\textwidth]{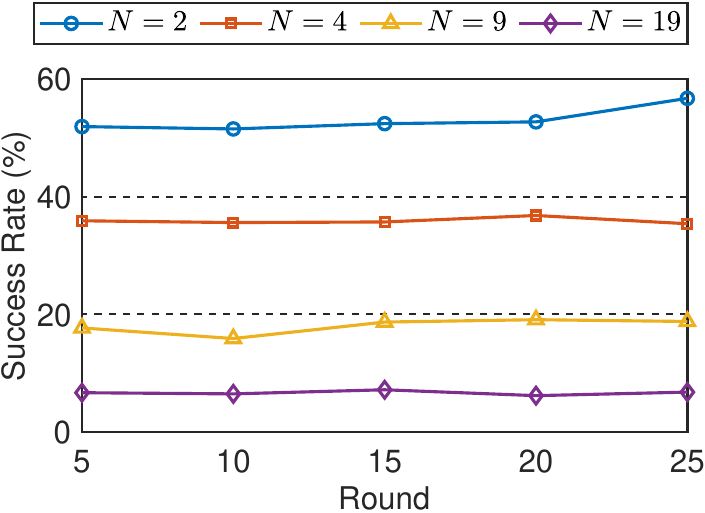}
    }
    \hspace{1em}
    \centering
    \subfigure[LLaMA3-8B.]{
    \label{fig:attack2}
        \includegraphics[width=0.28\textwidth]{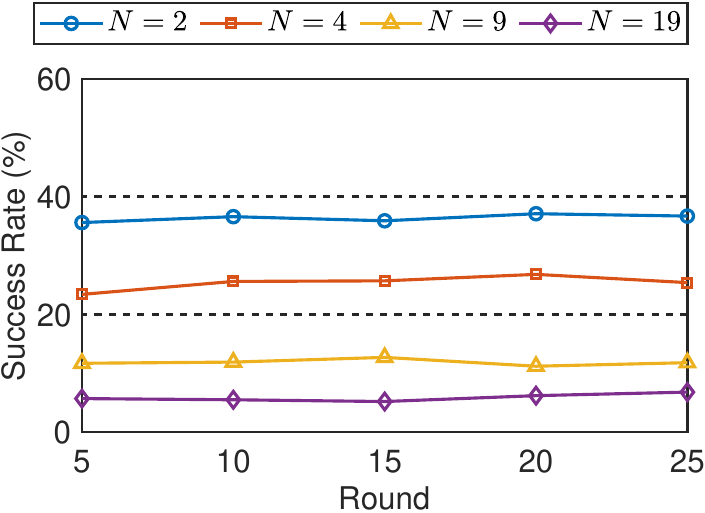}
    }
    \hspace{1em}
    \centering
    \subfigure[LLaMA3-13B.]{
    \label{fig:attack3}
    \includegraphics[width=0.28\textwidth]{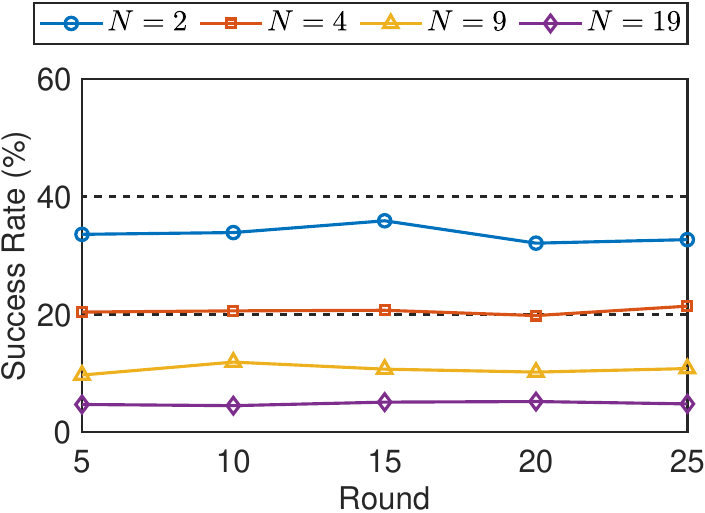}
    }
    \caption{Average attack success rate across query rounds for different values of $N$ and base models.}
    \label{fig:attack}
    \vspace{-1em}
\end{figure*}

\subsection{Privacy Protection (Q2)}
\label{sec:privacy}

\subsubsection{Qualitative risk analysis}

Before quantifying \textit{HomeLLaMA}'s privacy protection, it is important to first examine potential privacy risks qualitatively. In typical cloud-based smart home systems, users must register for API keys and transmit commands and device states to remote servers, exposing them to risks during data storage, network transmission, and inference, as detailed in Table \ref{tab:risk}. These include unauthorized access, interception, and inference attacks such as PII extraction, attribute inference, and activity monitoring. Unlike such systems, \textit{HomeLLaMA} operates locally without sharing user profiles or home configurations, thereby mitigating storage and transmission risks. Additionally, its \textit{PrivShield} module filters sensitive content before sending prompts to the cloud, preventing PII exposure. However, obfuscated prompts remain susceptible to activity monitoring via API traceability. The next section presents quantitative experiments to evaluate resilience against this residual threat.

\subsubsection{Quantitiave Analysis}

We focus on examining the in-home activity monitoring threat in \textit{HomeLLaMA}. During the use of \textit{PrivShield}, the real commands are obfuscated with \(N\) other SLM-generated adversarial commands before being sent to the cloud LLM for processing. 

\noindent \textbf{Threat Model.}
We assume the cloud LLM operates on the \textit{honest-but-curious} remote server \cite{r44}, where adversaries 
deliver the correct inference results but investigate all transmitted user queries. The goal is to identify the real query from the mixture using pretrained classifiers, thereby enabling real-time monitoring of users' in-home activities. 
Following the procedures in \cite{r121}, we launch the activity monitoring attack by instructing cloud GPT-4 using well-designed prompts shown in Fig. \ref{fig:attack prompt} to infer user in-home activities based on the continuously received commands. We then use \textit{attack success rate} \cite{r120} to assess the system's privacy level, where a higher rate indicates a greater threat to user privacy and reduced system protection.

\noindent \textbf{Results.}
\label{risk results}
We use the constructed \textit{DevFinder} as test user queries, setting the number of adversarial commands \(N\) to 2, 4, 9, and 19. We also vary the base SLMs in \textit{HomeLLaMA} to examine their impacts on the quality of adversarial command generation. The average attack accuracy results across query rounds obtained through extensive experiments are presented in Fig. \ref{fig:attack}, and we report the following key findings:

\begin{itemize}[leftmargin=9pt]
    \item \textit{PrivShield} effectively safeguards user privacy by maintaining significantly lower attack success rates compared to direct queries without its protection. As shown in Fig. \ref{fig:attack}, leveraging \textit{PrivShield} with different $N$ values and various base models results in a substantial reduction in attack accuracy, far below the 100\% success rate of direct queries. Furthermore, while an increase in query rounds allows adversaries to accumulate more information, as illustrated in all sub-figures, this does not enhance their ability to accurately infer user prompts.
    In fact, denote the average attack success rate of the \textit{PrivShield} as $\text{SR}_p$, and the overall attack success rate of \textit{HomeLLaMA} $\text{SR}_h$ should be
    \begin{equation}
        \text{SR}_h = \epsilon \cdot \text{SR}_p
    \end{equation}
    where $\epsilon$ represents \textit{PrivShield}'s frequency of use.
    During practical daily usage, the frequency $\epsilon$ will gradually become lower with user profiles being progressively constructed, as discussed in \S~\ref{sec:7.2}. Consequently, the overall privacy protection of \textit{HomeLLaMA} will strengthen over time, as users will increasingly rely on local operations without cloud LLM assistance.
    \item Privacy protection strengthens as the number of adversarial commands increases. These adversarial commands can \textbf{introduce noises in the text space}, making it harder for malicious attackers to identify the real queries.
    Moreover, generating more adversarial commands incurs higher latency and cost, users should be allowed to decide their preferred trade-off between privacy and performance.
    \item Privacy protection also benefits from the use of stronger SLMs. As demonstrated in Table \ref{tab:risk}, replacing base SLMs with larger and more powerful models significantly reduces attack accuracy. This is because \textbf{identifying real user queries becomes equivalent to distinguishing AI-generated text from the mixture}. Stronger SLMs are more adept at generating high-quality adversarial commands, further obscuring the real query from identification. However, deploying larger models may be impractical due to resource constraints, which will be discussed further in \S~\ref{sec:sensitivity}.
\end{itemize}

\noindent \textbf{Remarks.} Although adversaries may deploy advanced pretrained classifiers to distinguish user commands from obfuscated mixtures. In such cases, \textit{PrivShield} could be enhanced by strengthening query obfuscation (\eg, selecting a larger $N$) for stronger privacy protection according to users' requirements.



\subsection{System Cost (Q3)}

\subsubsection{Metrics}
We evaluate the system cost in terms of the following aspects: \textbf{1) Response latency:} We measure the time cost from the moment the user inputs a command to the generation of the final action plan as the response latency of each system. \textbf{2) Memory usage:} We measure the system overhead by tracking the GPU memory usage (in GB) via a Python package named memory-profiler \cite{r84} during usage.

\subsubsection{Baselines}
We keep the same baselines as selected in \S~\ref{sec:baselines}. Note that the first three systems (\ie, HomeGPT, Sasha, and SAGE) are based on cloud LLMs which cannot be directly accessed, while only TT explores the integration of SLMs into smart home assistants. Consequently, we only measure the memory usage of TT-Gemma, TT-Phi-2, and our proposed \textit{HomeLLaMA}.

\begin{figure}
    \centering
    \includegraphics[width=0.8\linewidth]{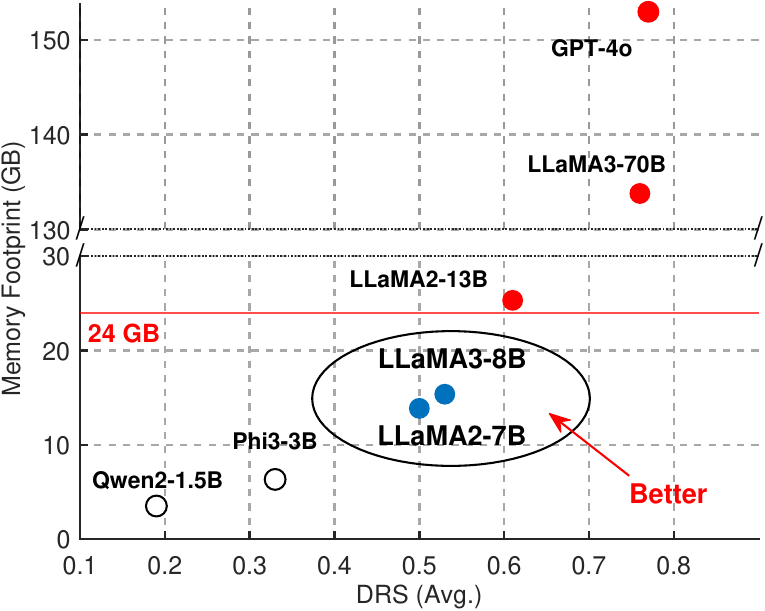}
        \caption{Impacts of different base models.}
        \label{fig:sen1}
        \vspace{-1em}
\end{figure}

\subsubsection{Results}
To measure the system overhead of \textit{HomeLLaMA} and the baselines, we input each test command in the \textit{DevFinder} benchmark into them and report the average response latency with its variance. As shown in Fig. \ref{fig:11a}, the cloud-based assistants exhibit relatively faster average response time (around 4.97 seconds) compared to the local-based assistants, primarily due to the performance optimizations of OpenAI services \cite{r20}. However, cloud-LLM-based systems are highly susceptible to network conditions and server stability, leading to significant variance in response latencies, which can negatively impact user experience \cite{r83}. In contrast, \textit{HomeLLaMA} exhibits less variance in response latency, albeit with a slightly longer response time. We also track the maximum GPU memory usage of the two local SLM-based systems (\ie, TT and \textit{HomeLLaMA}) when adopting different quantization precisions (\ie, fp16, int8, and int4) during inference. As shown in Fig. \ref{fig:11b}, the maximum GPU memory requirement for these systems remains under 16 GB, which is affordable and manageable for a typical household\footnote{An NVIDIA RTX 4070 Ti SUPER GPU (16 GB) costs around \$840.}. 


\subsection{Sensitivity Analysis (Q4)}
\label{sec:sensitivity}

\subsubsection{Different base models}
We evaluate the impact of base models by deploying Qwen2-1.5B~\cite{r77}, Phi3-3B~\cite{r78}, LLaMA2-7B/13B~\cite{r79}, LLaMA3-8B~\cite{r16}, and LLaMA3-70B~\cite{r79} in \textit{HomeLLaMA}. Given the test inputs from \textit{DevFinder}, we record their average \textit{DRS} and GPU memory usage. As shown in Fig.~\ref{fig:sen1}, models fall into three groups: (1) lightweight models (white) such as Qwen2-1.5B and Phi3-3B offer minimal memory usage (${\sim}$8GB) but poor performance; (2) high-end models (red) like LLaMA2-13B, LLaMA3-70B, and GPT-4o yield the best \textit{DRS} but require ${\geq}$24GB GPU memory; (3) mid-range models (blue) such as LLaMA2-7B and LLaMA3-8B balance performance and cost. Thus, we choose LLaMA3-8B as the default base model.

\subsubsection{Different $\alpha$ during augmentation}
To study the effect of the augmentation threshold $\alpha$ (Eq.~\ref{eq:alpha}), we vary it from 0.1 to 0.9 and fine-tune LLaMA3-8B, Phi3-3B, and LLaMA2-7B on corresponding augmented datasets. As shown in Fig.~\ref{fig:13a}, increasing $\alpha$ improves average \textit{DRS} by promoting data diversity. However, the gain saturates at higher values, while training cost rises due to dataset expansion. We therefore set $\alpha=0.7$ by default.

\subsubsection{Different $\beta$ during profile updating}
\begin{figure}
        \centering
        \subfigure[Distinct $\alpha$ for the inspector.]{
            \includegraphics[width=0.45\linewidth]{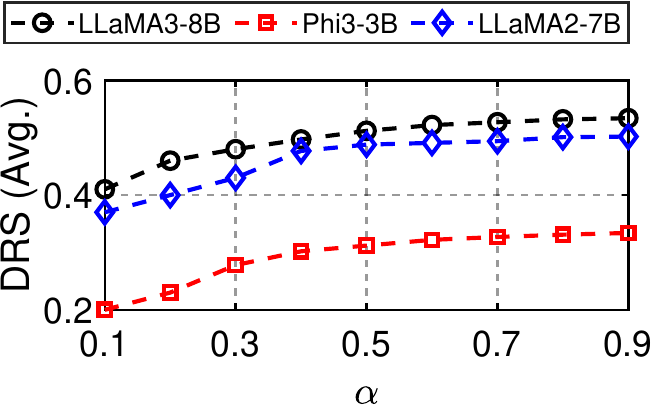}
            \label{fig:13a}
        } 
        \centering
        \subfigure[Distinct $\beta$ for merging profiles.]{
            \includegraphics[width=0.45\linewidth]{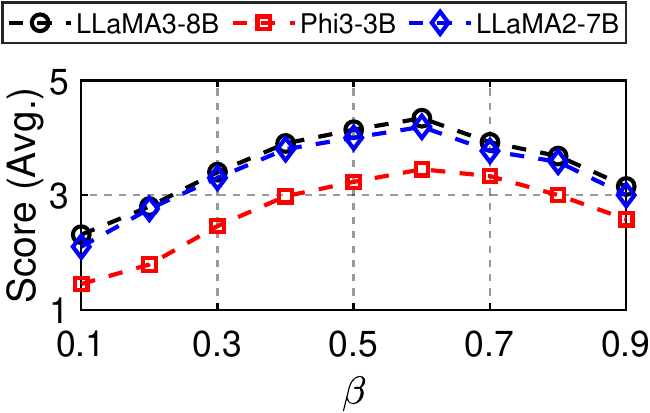}
            \label{fig:13b}
        }
        \caption{Impacts of selecting different thresholds $\alpha$ and $\beta$.}
        \label{fig:thres}
        \vspace{-1em}
\end{figure}

The profile update threshold $\beta$ (Eq.~\ref{eq:beta}) determines when to save a new user profile. Using 10 participants, we evaluate personalization ratings (1–5) under varying $\beta$ from 0.1 to 0.9. As shown in Fig.~\ref{fig:13b}, higher $\beta$ initially improves satisfaction by preventing premature profile merging, but overly high values lead to redundant entries, impairing retrieval. Satisfaction peaks around $\beta=0.6$, which we adopt as the default.

%% file: sections/07_user_study.tex
We conduct user studies to gather user feedback. The study can be divided into two parts: an \textit{online survey} and an \textit{onsite interview}, aiming to answer those questions:

\begin{itemize}[leftmargin=9pt]
    \item \textbf{Q5} - \textit{User Privacy Confidence:} Does \textit{HomeLLaMA} lift user-perceived privacy confidence?
    \item \textbf{Q6} - \textit{User Satisfaction:} How do users feel about the overall service quality of \textit{HomeLLaMA}?
    \item \textbf{Q7} - \textit{Long-term Personalization:} Is \textit{HomeLLaMA} capable of adapting to users' preferences continuously? 
\end{itemize}

\subsection{Online Survey: Cold-start Evaluation (Q5 \& Q6)}
\label{sec:online}


\subsubsection{Preparation works}
\label{sec:7.1}

We select two representative baselines: a cloud-based assistant SAGE and a local-based assistant TT-Gemma. For each scenario, we select a test command from \textit{DevFinder} and input them into systems to generate initial action plans. Subsequently, we manually select the "Advice" option (\S~\ref{sec:4.2.1}) and provide the feedback to all systems. These systems then perform preference learning, if applicable, and regenerate action plans which are recorded for further analysis. 
To ensure fairness, the order of system presentation is randomized before being rated by participants.

\begin{figure*}[t]
    \centering
    \subfigure[Overall ratings (survey).]{
        \includegraphics[width=0.23\textwidth]{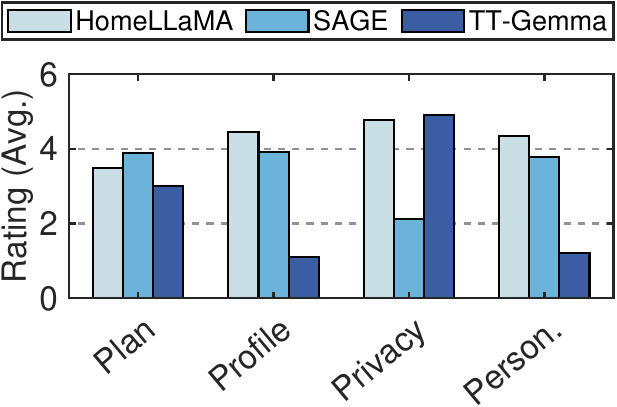}
        \label{fig:study1}
    }
    \hfill
    \subfigure[Long-term personalization.]{
        \includegraphics[width=0.23\textwidth]{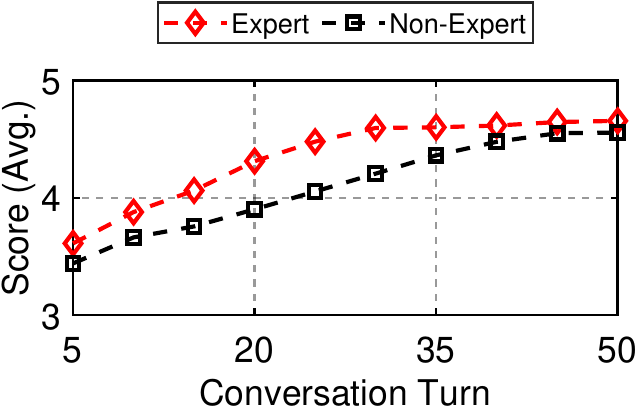}
        \label{fig:21a}
    }
    \hfill
    \subfigure[Ease of use.]{
        \includegraphics[width=0.23\textwidth]{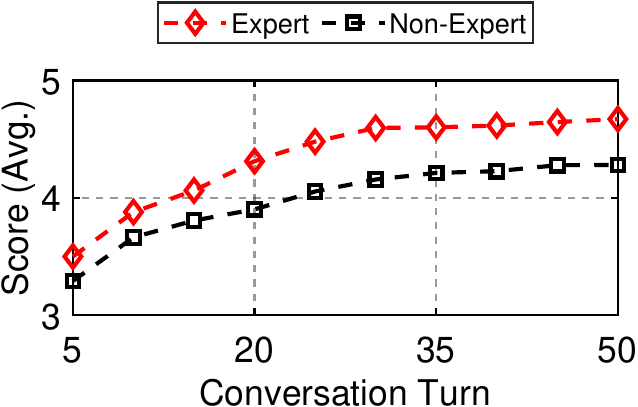}
        \label{fig:21b}
    }
    \hfill
    \subfigure[Use frequency of \textit{PrivShield}.]{
        \includegraphics[width=0.23\textwidth]{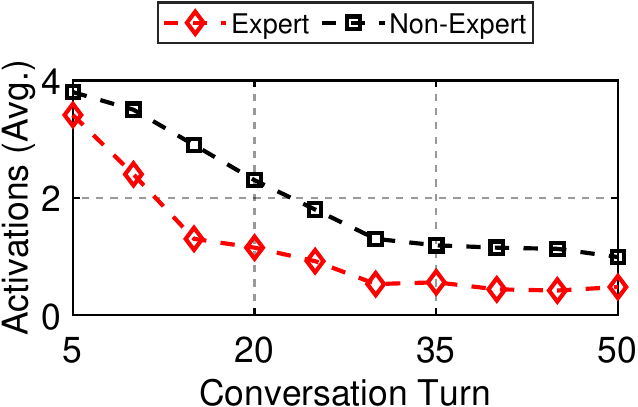}
        \label{fig:21c}
    }
    \caption{User study results, including (a) online survey ratings, and (b)--(d) show average interview results.}
    \label{fig:per_combined}
    \vspace{-1em}
\end{figure*}

\subsubsection{Participants}
We created an online survey using Microsoft Forms\footnote{\url{https://forms.office.com/}} and distributed it via emails and social media to recruit participants. Before participating, all respondents were presented with an informed consent form outlining the purpose of the study, data handling procedures, and their rights as participants. The study protocol was reviewed and approved by our institution’s ethics review board (IRB). Participants were informed that their responses would be anonymized and used solely for academic research purposes. Participation was entirely voluntary, and no monetary compensation was provided. After 12 days, we received \textbf{100 responses} from volunteers in total, and the data of participants shows a relatively balanced distribution of participants in terms of gender, age, educational background, English proficiency, and familiarity with smart home assistants.

\subsubsection{Survey design}


The survey evaluated smart home systems from the following four metrics: \textcircled{1} \textbf{General Plan Satisfaction}, assessing satisfaction with the quality of initial action plans irrespective of personalization; \textcircled{2} \textbf{User Profile Correctness}, measuring how accurately the generated user profile reflects personal preferences and behaviors during the current interaction; \textcircled{3} \textbf{Personalization Fit Score}, evaluating how well the responses and recommendations align with historical feedback within the interaction round; and \textcircled{4} \textbf{Privacy Assurance Score}, gauging participants' confidence in the system's ability to safeguard personal data and maintain privacy. Participants were asked to rate these metrics on a Likert scale \cite{r124} from 1 (\eg, not at all) to 5 (\eg, completely).

\subsubsection{Overall results}

Fig.~\ref{fig:study1} visualizes the overall rating results, and key observations of each aspect include:
1) Participants expressed high satisfaction (close to cloud-based assistant SAGE) with the quality of services provided by \textit{HomeLLaMA}, reflecting its ability to deliver effective and reliable action plans tailored to user needs. This may be attributed to the tailored enhancement method of SLMs proposed in this paper.
2) \textit{HomeLLaMA} excelled in delivering personalized responses and generating precise user profiles compared with other baselines, achieving the highest average score of around 4.35 points. This may be attributed to the \textit{User Preference Learning} module, which accurately characterizes well-structured user profiles to adapt plans effectively.
3) \textit{HomeLLaMA} received high ratings for its privacy-preserving features, surpassing cloud-based solutions and approaching the level of the fully local system, TT-Gemma. This demonstrates its ability to enhance user-perceived privacy by operating locally and minimizing data transmission to the cloud. Additionally, the local-cloud collaboration paradigm, supported by the designed \textit{PrivShield}, boosts users' confidence in privacy, thereby improving the overall usability of the system.

\subsection{Onsite Interview: Long-term Evaluation (Q7)}
\label{sec:7.2}

To evaluate the long-term performance of \textit{HomeLLaMA}, we deploy the system locally for conducting an onsite interview. The interview lasts for 25 days with 50 conversation turns, and the actual usage time of volunteers is approximately 45.6 minutes on average. Each turn represents a complete user-assistant-cloud interaction (Fig.~\ref{fig:flow}), starting with the user command and ending with the final action plan. 

\subsubsection{Procedures}

We deployed \textit{HomeLLaMA} on a laboratory PC using the configuration in \S~\ref{sec:implement}. From 100 survey respondents described in \S~\ref{sec:7.1}, 10 participants were invited and divided into two groups: 5 experts and 5 non-experts, based on their familiarity with smart homes. Prior to interviews, all participants provided informed consent and received compensation in the form of supermarket coupons valued at approximately \$10. Participants were first introduced to the interaction workflow—including the accept, advise, and reject options—to ensure familiarity. Additionally, two evaluation metrics were explained to them before beginning the interaction: \textcircled{1} \textbf{Long-term Personalization} assesses how effectively the system infers and retains user preferences over time. \textcircled{2} \textbf{Ease of Use} evaluates how efficiently the system minimizes user efforts for delivering satisfactory responses as usage continues. Participants were then invited to freely interact with \textit{HomeLLaMA}, issuing smart home commands in natural language through UI without constraints. After every five conversation turns, they rated the system using the two predefined metrics on a 5-point scale. Throughout the session, we recorded all evaluation scores and, every five turns, also tracked the number of times \textit{PrivShield} was activated for cloud assistance. Finally, the average results for each group were computed and analyzed separately.

\subsubsection{Results}

As illustrated in Fig. \ref{fig:per_combined}, both expert and non-expert participants show a steady increase in ratings for \textit{HomeLLaMA} across personalization and ease of use as the number of conversation turns grows, reflecting the system's ability to adapt through dynamically maintained user profiles. Experts tend to assign higher scores earlier, likely due to their clearer articulation of preferences, which accelerates profile quality and system adaptation. A consistent gap remains in ease-of-use ratings, with experts maintaining an advantage of about 0.35 points by the 50th turn, though both groups converge above 4.2, indicating strong usability. Additionally, the activation frequency of \textit{PrivShield} declines for both groups, approaching zero by the 50th turn, highlighting \textit{HomeLLaMA}’s reduced reliance on cloud-based support as it better internalizes user preferences, thereby enhancing privacy protection.

%% file: sections/11_conclusion.tex
This paper presents an on-device smart home assistant that balances privacy and performance for users. The designed \textit{HomeLLaMA} comprises three technical modules: \textit{Local SLM Enhancement}, \textit{Multi-Parity Interaction}, and \textit{User Preference Learning}, enabling seamless and privacy-enhanced interactions involving user-in-the-loop. We construct the \textit{DevFinder} benchmark to assess the quality of the generated responses. Comprehensive user studies demonstrate that \textit{HomeLLaMA} delivers satisfactory plans with enhanced personalization, while alleviating privacy concerns. Additionally, extensive quantitative experiments verify the effectiveness of \textit{HomeLLaMA} in enhancing user privacy in data storage, network transmission, and inference-related (\eg, PII extraction) attacks.